\documentclass{emulateapj}
\usepackage{epsfig}
\usepackage{natbib}
\slugcomment{To appear in the Astrophysical Journal}

\shorttitle{A $Spitzer$ Study of IC 405}
\shortauthors{France et al.}

\begin{document}


\title{A Cometary Bow Shock and Mid-Infrared Emission Variations  
Revealed in $Spitzer$ Observations of HD 34078 and IC 405\altaffilmark{1}}


\author{Kevin France\altaffilmark{2}, Stephan R. McCandliss, and Roxana E. Lupu}
\affil{Department of Physics and Astronomy, Johns Hopkins University,\\
	3400 N. Charles Street, Baltimore, MD 21218}



    

\altaffiltext{1}{Based in part on observations made with the the $Spitzer$~$Space$~$Telescope$, 
which is operated by the Jet Propulsion Laboratory, California Institute of Technology, 
under a contract with NASA.}

\altaffiltext{2}{Current Address: Canadian Institute for Theoretical Astrophysics, University of Toronto,
    60 St. George Street, Toronto, ON M5S 3H8}

\received{2006 August 31}
\revised{2006 October 18}
\accepted{2006 October 27}



\begin{abstract}
We present new infrared observations of the emission/reflection nebula
IC 405 obtained with the $Spitzer$~$Space$~$Telescope$.  Infrared 
images in the four IRAC bands (3.6, 4.5, 5.8, and 8.0 $\mu$m) and
two MIPS bands (24 and 70 $\mu$m) are complemented by 
IRS spectroscopy (5~--~30 $\mu$m) of two nebular filaments. 
The IRAC (8.0 $\mu$m) and MIPS imaging shows evidence of a bow shock associated 
with the runaway O9.5V star, HD 34078, created by the interaction between the star and nebular material.  
The ratio of emission at 24 to 70 $\mu$m is higher 
in the immediate vicinity of HD 34078 than in the outer filaments, providing
evidence for elevated dust temperatures (T$_{d}$~$\gtrsim$~90 K) in the shock region. 
The nebular imaging reveals that the morphology is 
band dependent, with varying contributions
from aromatic emission features, H$_{2}$, and dust emission. 
Nebular spectroscopy is used to quantify these contributions,  
showing several aromatic emission bands between 6~--~14 $\mu$m, 
the S(5), S(3), S(2), and S(1) pure rotational emission lines of 
H$_{2}$, and atomic 
fine structure lines of Ne, S, and Ar.
The low-dispersion spectra provide constraints on the ionization 
state of the large molecules responsible for the aromatic infrared features.
H$_{2}$ rotational temperatures of the two bright nebular filaments
are determined from the observed line strengths.
An average T(H$_{2}$)~$\sim$~400 K is inferred, with evidence for additional
non-uniform excitation by UV photons in the intense radiation field 
of HD 34078.  The photoexcitation
hypothesis is supported by direct measurement of the far-UV 
H$_{2}$ fluorescence spectrum, obtained with $FUSE$.
\end{abstract}



\section{Introduction}

IC 405 is an emission/reflection nebula in Auriga~\citep{herbig58,herbig99}.
It is illuminated by the O9.5V star HD 34078 (AE Aur, V~=~6.0), 
a high proper motion ($\approx$ 40 mas/yr; Blaauw \& Morgan 1954)
runaway star ejected from the Orion star-forming region in a binary-binary
interaction approximately 2.5 million years ago~\citep{bagnuolo01}. 
The star and the optical nebula are roughly cospatial at 
present~\citep{herbig58,france04,boisse05}, at a distance of 450~pc.     
The sightline to HD 34078 is considerably reddened ($E(B~-~V)$~=~0.53;
Cardelli et al. 1989)~\nocite{ccm}, however, the extended nebular regions
appear to be free of foreground attenuation~\citep{france04}.

The large transverse velocity of the star indicates that it has only 
recently encountered the nebular material with which it now interacts.
The present geometric configuration has made the HD 34078 sightline 
a target for studies of interstellar absorption line variation.
\citet{herbig99} combines fifty years of atomic and molecular line 
observations in the violet-blue ($\lambda$3400~--~4900 \AA) to set an upper limit on the length scale 
of substructure in the ISM of 1450 AU.  More recently, variations 
in the column density of interstellar CH and CN (10~--~20~\%) have been 
observed on this sightline, setting a smaller upper limit on the 
substructure size ($\leq$ 200 AU) towards HD 34078~\citep{rollinde03}.  A
similar study~\citep{boisse05} using $Far$ $Ultraviolet$ 
$Spectroscopic$ $Explorer$ ($FUSE$) observations has set an upper limit 
on the molecular hydrogen (H$_{2}$) column variation of $\leq$~5\%, 
constraining the scale to a few tens of AU.  They also find evidence 
for interaction between HD 34078 and the surrounding nebula.  Detections
of excited molecular hydrogen absorption lines (measurements up to
$v$~=~0, $J$~=~11 and tentative detections up to the $v$~=~2 level) are 
evidence for fluorescent pumping by the local stellar UV field~\citep{boisse05}.
Additional observations of the interaction between the UV radiation 
field of HD 34078 and the nebular gas and dust are presented by~\citet{france04}.
They present a spectroscopic study of far-UV dust scattering and 
fluorescent H$_{2}$ emission lines in the nebula.

Evidence for interaction between the star and nebular dust grains has also been
detected by $IRAS$~\citep{vanburen95}.  The high proper motion of HD 34078 makes it a 
candidate host for a bow shock.  As a massive star moves supersonically through an 
ambient interstellar medium, stellar radiation pressure and/or strong stellar winds can produce a   
comet-shaped, compact \ion{H}{2} region surrounded by shocked molecular 
gas~\citep{vanburen88,vanburen90}.  A 60 $\mu$m excess observed in 
$IRAS$ maps~\citep{vanburen95,noriega97} identifies IC 405 as a potential bow shock region.  

In this paper, we present the first direct imaging of the bow shock around HD 34078
with the $Spitzer$ $Space$ $Telescope$-IRAC and MIPS.  We also use MIPS 
data to study the nebular dust emission, characterize the spectroscopic properties
of aromatic emission features (AEFs) and H$_{2}$
using nebular observations made with the $Spitzer$-IRS, 
and put these new observations in context with existing studies of IC 405.
\S1.1 provides background information
on the diagnostic properties of AEFs, H$_{2}$, and 
dust in diffuse nebulae.  In \S2, we describe 
the new $Spitzer$ data, and introduce supporting
far-UV spectroscopy from $FUSE$.  An analysis of the 
infrared (IR) imaging and spectroscopy is given in \S3.
In \S4, we discuss the structure of the bow shock region, 
present determinations of the H$_{2}$ rotational and dust temperatures,
and analyze spatial variations in AEF carrier ionization
across the nebular filaments.  
Our findings and a summary of the observations are presented in \S5.

\begin{figure*}
\hspace{+0.0in}
\includegraphics[width=4.0in]{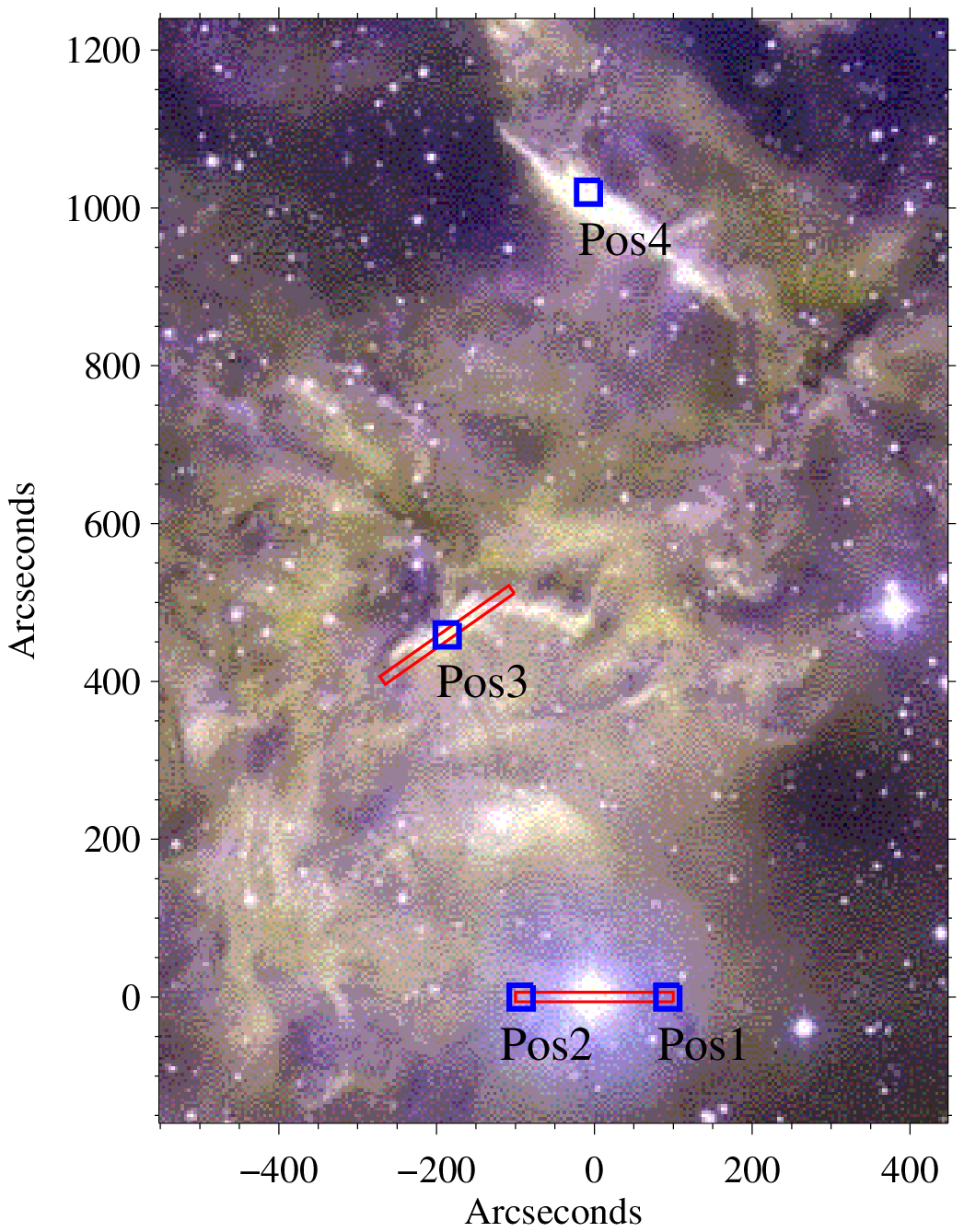}
\hspace{-0.3in}
\includegraphics[width=4.0in]{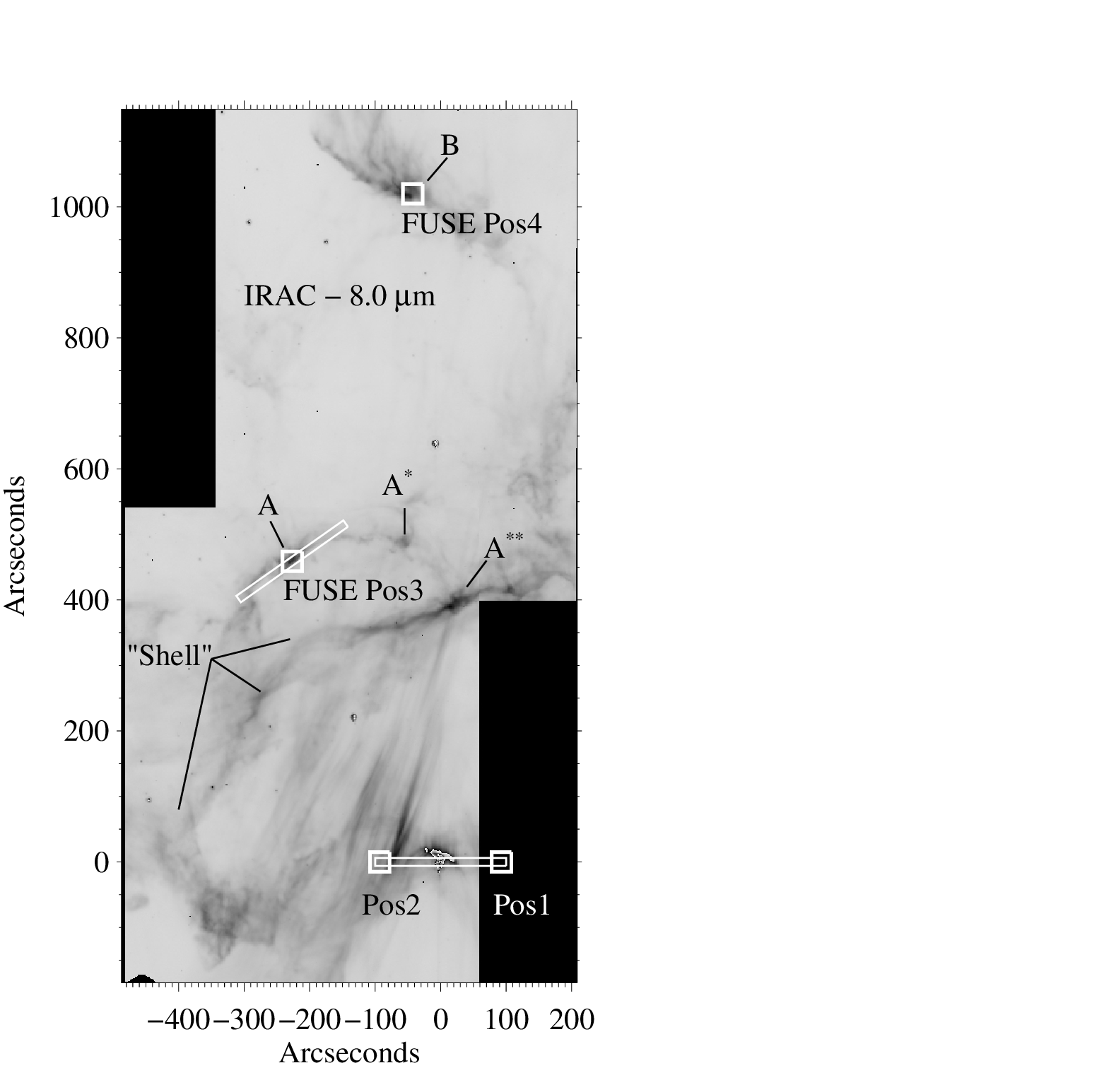}
\caption{\label{irac405} 
$Left$: Optical image of IC 405 from~\citet{france04}.
$Right$: Post-BCD 8.0~$\mu$m IRAC images of IC 405, 
scaled to highlight the rich nebular structure.
The origin is the position of the illuminating, runaway O9.5 star, HD 34078.
The main nebular regions studied in this work are labeled
A and B.  White overlays represent far-UV 
spectrograph apertures (the squares are the
$FUSE$ LWRS slit and the rectangles represent a rocket-borne spectrograph; France et al. 2004).
At the cores of the bright filaments, discrete features 
contribute $\leq$ 40\% of the nebular emission at 8 $\mu$m.  
}
\end{figure*}

\subsection{AEFs, H$_{2}$, and Dust Emission in Photodissociation Regions}
AEFs have been found 
to be ubiquitous in the mid-IR spectra (3~--~20~$\mu$m) of reflection
nebulae and \ion{H}{2} regions (see van Dishoeck 2004 for a recent review).
\nocite{vand04,werner04,peeters04}
AEFs are seen as broad bands in the 
$ISO$-SWS/$ISO$-CAM-CVF and $Spitzer$-IRS wavelength
ranges, with lines complexes centered on 3.3, 6.2, 7.7, 8.6, 11.2, and
12.7~$\mu$m.  These features are usually attributed to a superposition 
of C~--~C and C~--~H vibrational modes in a distribution of polycyclic aromatic 
hydrocarbons (PAHs) 
consisting of tens to hundreds of C-atoms~\citep{schutte93,vand04}. 
However, discrepancies between astronomical emission features and laboratory PAH spectra 
exist, preventing a conclusive identification of the exact carrier of the AEFs 
(Gordon et al. 2006 and references therein).~\nocite{gordon06}
The carrier population that gives rise to the observed mid-IR bands (also
called Aromatic Infrared Bands; Verstraete et al., 
2001)\nocite{verstraete01} spans a range of 
sizes and ionization states.  The size distribution is thought 
to be regulated by the hardness of the illuminating radiation 
field~\citep{verstraete01} while the aromatic molecule ionization balance is
controlled by the strength of the ultraviolet (UV) radiation field
and the electron density~\citep{bakes01,bregman05}.  The relative strengths and 
spectral profiles of the major AEFs reflect the physical state
of the carrier population in a given environment.


Variations in the spectral features of the AEFs can be used to quantify
the response of the carrier molecules to 
different radiation and density environments~\citep{bakes01}.  
These variations can be studied by comparing a sample of nebulae 
~\citep{verstraete01}, at different locations within a single object~\citep{verstraete96,hony01}, 
or in the diffuse interstellar medium~\citep{boulanger00}.
The C~--~H emission features at
3.3 and 11.2~$\mu$m are found to be well correlated over a range of
physical conditions, however the relative strengths of the C~--~H and
C~--~C features are observed to vary with environment~\citep{hony01,vand04}.
The ratio of a C~--~C  to a C~--~H feature (6.2/11.2 $\mu$m for example)
reflects the ionization state as C~--~C band strengths are
stronger in ionized aromatic carriers while C~--~H bands 
display the opposite behavior~\citep{uchida00}.  
The increase in the 6.2 $\mu$m C~--~C feature 
(and more generally, the 6.2, 7.7, and 8.6 ~$\mu$m bands) is due to the dipole moment
produced by a change in charge distribution during ionization~\citep{peeters02}.
The carriers of the AEFs fall in the middle of the
molecular mass distribution in reflection nebulae and \ion{H}{2} 
regions (which we will group under the broad heading of
photodissociation regions; PDRs).  

\begin{figure}
\begin{center}
\hspace{+0.0in}
\includegraphics[angle=90,width=7.5in]{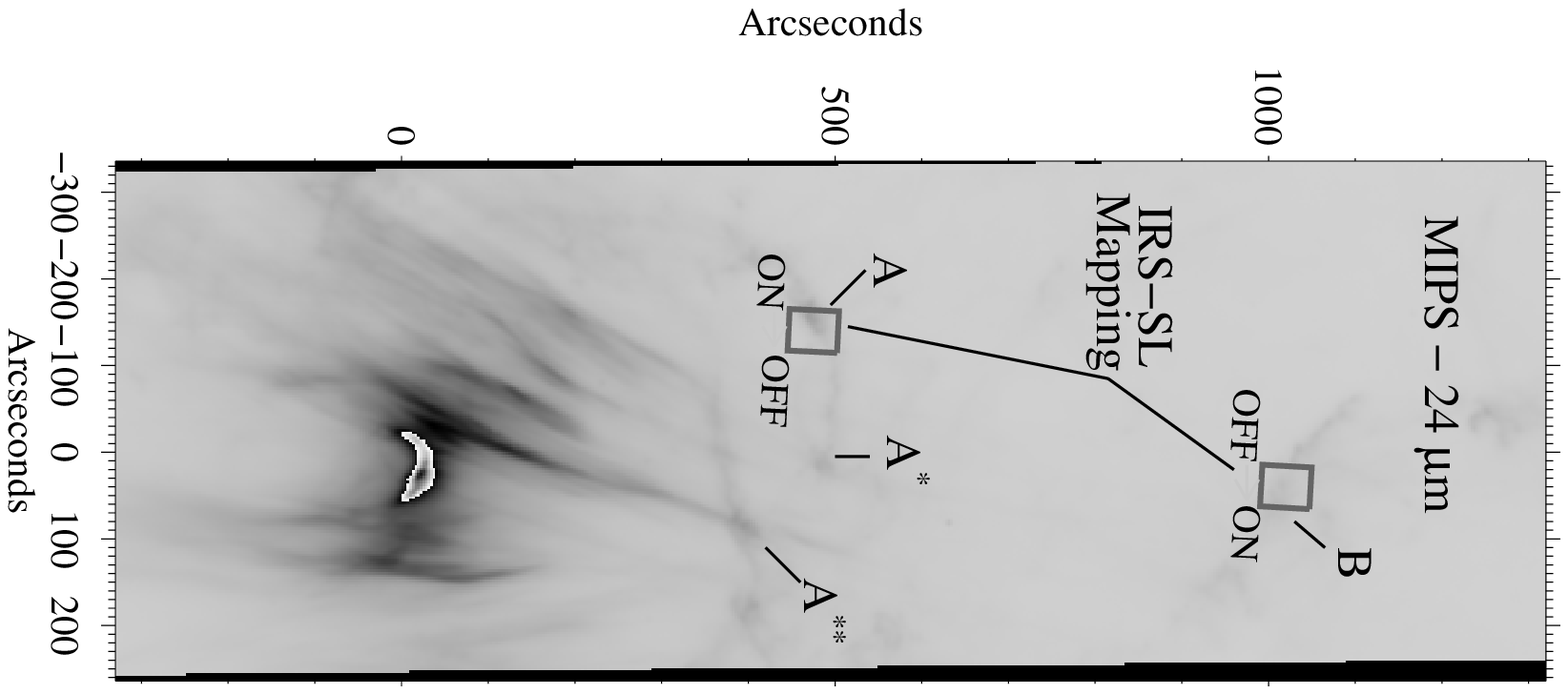}
\caption{\label{mips24} Post-BCD 24 $\mu$m map of IC 405.
The region near HD 34078 (0,0) shows the wake-like morphology, suggesting the
presence of a bow shock as the star traverses the ambient nebular material.
The bow shock region at 8 and 24 $\mu$m can be seen enlarged
in Figures 4 and 5.  The overlays on Filament A and B represent the 
IRS-SL scan regions, with the ON/OFF convention used in Figures 10 and 12
indicated. (The IRS-SL scans are shown in more detail in Figures 8 and 9.)}
\end{center}
\end{figure}

H$_{2}$, residing at the lower
limit of this distribution, makes up the majority of the 
molecular mass in these objects~\citep{pdrs,shaw05}.  H$_{2}$ 
is seen to emit from far-UV to the mid-IR wavelengths, 
excited by both UV photons (i.e. - fluorescence) and thermal processes, such as collisions and shocks.  
Detection of the UV emission lines of H$_{2}$ is one clear indication of 
photoexcitation. H$_{2}$ dissociates in environments with temperatures 
corresponding to the excited electronic levels that give rise to the UV lines~\citep{shull82}. 
The emission nebula IC 63 is the prototypical object for PDR studies of fluorescent H$_{2}$.  
The far-UV spectrum has been well-studied~\citep{witt89,hurwitz98,france05a},  
confirming stellar continuum photons as the excitation source of the observed emission.
Far-UV H$_{2}$ emission lines in IC 405 have been observed~\citep{france04}, 
and these observations will be discussed below.

Near-IR observations of the rovibrational emission lines of H$_{2}$ can also be used to constrain 
the molecular excitation mechanism.  Collisional processes can populate the lowest
excited vibrational states of the molecule ($v$~=~1 and 2), but the fluorescent 
cascade associated with the electronic excitation of H$_{2}$ by UV photons is the only
mechanism (ignoring excitation by non-thermal electrons)
available to populate higher vibrational states ($v$~$>$~2) as the molecules
dissociate at kinetic temperatures proportional to these higher vibrational levels~\citep{shull82}.
Ratios of these excited vibrational states can be compared with model predictions
to determine the excitation mechanism~\citep{black87}.   
Near-IR line ratios indicate UV excitation in IC 63~\citep{luhman97}, 
the bright reflection nebulae NGC 2023 and NGC 
7023~\citep{martini99,takami00}, and on larger scales in molecular clouds~\citep{luhman96}.
Collisions contribute to the rovibrational excitation of H$_{2}$ in 
dense molecular clouds~\citep{kristensen03} and regions where shocks play 
an important role, such as planetary nebulae~\citep{kastner96,lupu06}.  
The pure rotational emission lines of H$_{2}$,   
accessible only in the mid-IR, 
give diagnostic information about the molecular gas temperature in PDRs.  
Ratios of these lines have been used to derive temperatures in 
star-forming regions~\citep{rosenthal00,allers05}, PDRs~\citep{thi99,habart04}, 
and external galaxies~\citep{valentijn99,armus04}.

\begin{figure}
\begin{center}
\hspace{+0.0in}
\includegraphics[width=7.5in]{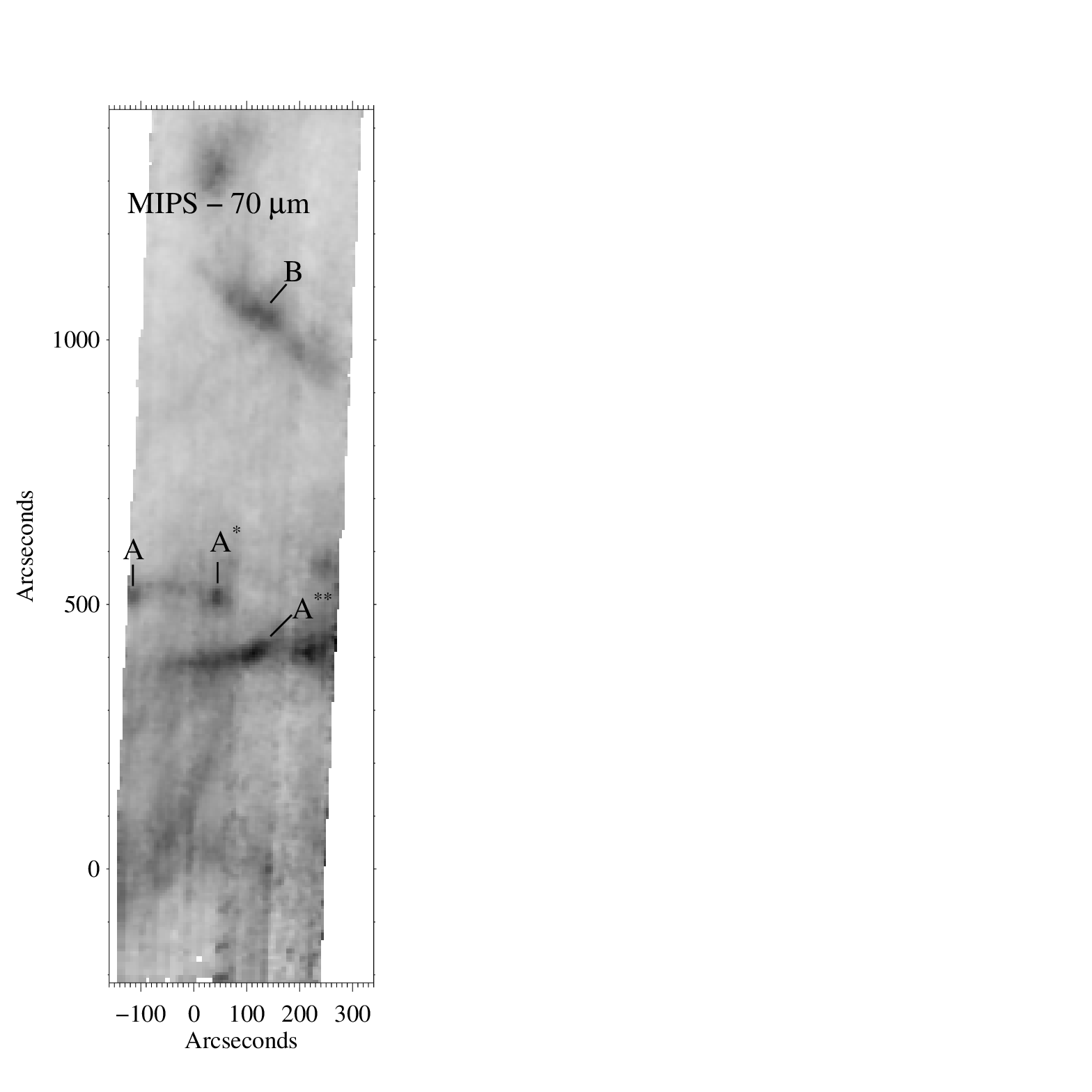}
\caption{\label{mips70} Post-BCD 70 $\mu$m scan of IC 405.
The position of HD 34078 is at the origin.  The Filament A$^{**}$ region is the strongest 
feature in the 70 $\mu$m data. }
\end{center}
\end{figure}

Dust grains populate the upper end of the molecular mass distribution in PDRs.
Grains are observed through their thermal emission in the mid/far-IR (Schnee et al. 2005 and
references therein) and in the attenuation and scattering of ultraviolet 
and visible photons~\citep{ccm,witt93,burgh02,draine03}.~\nocite{schnee05}
$IRAS$ observations of PDRs find that grain emission is spatially correlated with 
near-IR H$_{2}$ emission~\citep{luhman96}.  \citet{jansen94,jansen95}
present observations of  IC 63, where dust emission is coincident with the region of 
peak H$_{2}$ emission (at UV and IR wavelengths) as well as CO and other molecules. 
This correlation would be expected as dust grains are thought to be the 
primary site of H$_{2}$ formation in the interstellar medium~\citep{cazaux02,cazaux04}.
Additionally, photoelectric emission from small grains is an important 
source of heating in PDRs~\citep{abel05}.  
\citet{habart04} provide observational evidence for a correlation between AEFs and 
H$_{2}$, which they propose is related to H$_{2}$ formation on PAHs.  
With the wealth of observational evidence that these molecules are 
interrelated in PDRs, it follows naturally to study this group as a whole.


In this work, we further address 
the physical conditions of the AEF carriers and their relationship 
with H$_{2}$ and dust grains by presenting new
observations of the emission/reflection nebula IC 405 made with $Spitzer$.  
We use a combination of imaging and 
spectroscopy to constrain the variation in aromatic carrier properties with
changing radiation and dust environment through the nebula.  
Our aim is to present new IR observational characteristics
of large molecules, dust grains, and H$_{2}$ in the unique PDR environment
of IC 405, for comparison with previous far-UV observations. 


\section{Observations}

\subsection{$Spitzer$}
$Spitzer$ observations of IC 405 (Program ID 20434)
were made at positions coincident with the UV 
data presented in~\citet{france04}, where H$_{2}$ fluorescence
was observed above the level of dust scattered light 
($FUSE$ Pos3: RA~=~5$^{h}$16$^{m}$30.21$^{s}$,
$\delta$~=~+34$^{\circ}$24$^{'}$56.4$^{''}$ and $FUSE$ Pos4:
RA~=~5$^{h}$16$^{m}$18.64$^{s}$,
$\delta$~=~+34$^{\circ}$32$^{'}$25.2$^{''}$, J2000; see also \S2.2).
The Infrared Array Camera (IRAC) observed the nebula in mapping mode on 2005 September
23, with an exposure time of 12 seconds per frame,  
for a total integration time approximately 825 seconds.  Details of the
IRAC instrument can be found in~\citet{irac04}.  
The IRAC pointings covered the northern nebula in 
all four bands (r16033792), with the 4.5/8.0~$\mu$m channel acquiring
data of HD 34078 and the associated bow shock (r16033280, Figure~1).
The images were reduced with the data processing pipeline (S14.0.0)
provided by the $Spitzer$~$Science$~$Center$ (SSC). 
Analysis was performed on the post-Basic Calibrated Data
(BCD) products.

\begin{figure}
\begin{center}
\hspace{+0.0in}
\includegraphics[width=3.5in]{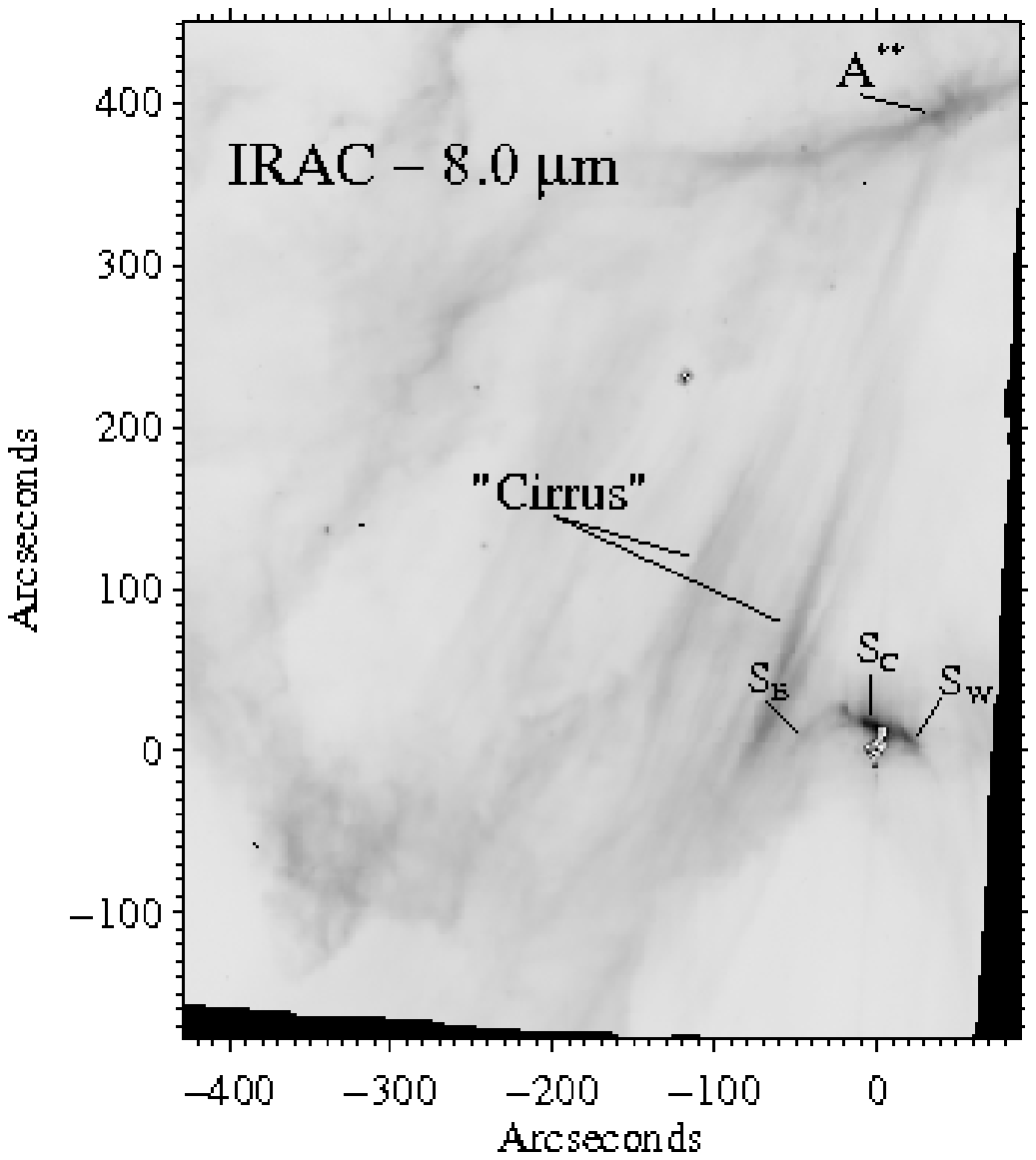}
\caption{\label{irac8bow} IRAC 8.0 $\mu$m blow up of the bow shock region, 
scaled to illustrate the shock structure.  The shock subregions are 
labeled Shock-East (S$_{E}$), Shock-Center (S$_{C}$), and Shock-West (S$_{W}$).
Photometry of the subregions is given in 
Table 1. 
}
\end{center}
\end{figure}

\begin{figure}
\begin{center}
\hspace{+0.0in}
\includegraphics[width=3.5in]{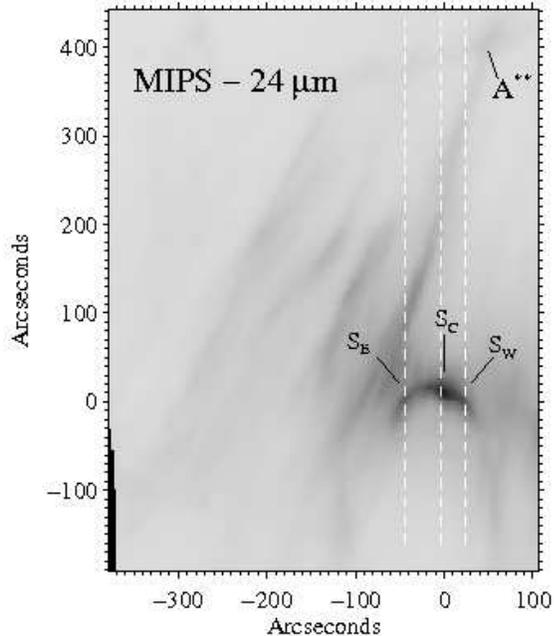}
\caption{\label{mips24bow} MIPS 24 $\mu$m blow up of the shock region, 
scaled to illustrate the shock structure.  The stellar position is at the origin.
The inner 10~--~25\arcsec\ near HD 34078 in the direction of stellar motion 
(North, +$y$) is saturated at 24 $\mu$m,  indicative of the high temperature of the 
grains in the shock region.  The dashed lines indicate the location and direction of 
the spatial profiles presented in Figure 7.
24 $\mu$m photometry of the shock region is given in Table 1 
}
\end{center}
\end{figure}

\begin{figure}
\begin{center}
\hspace{+0.0in}
\includegraphics[width=3.5in]{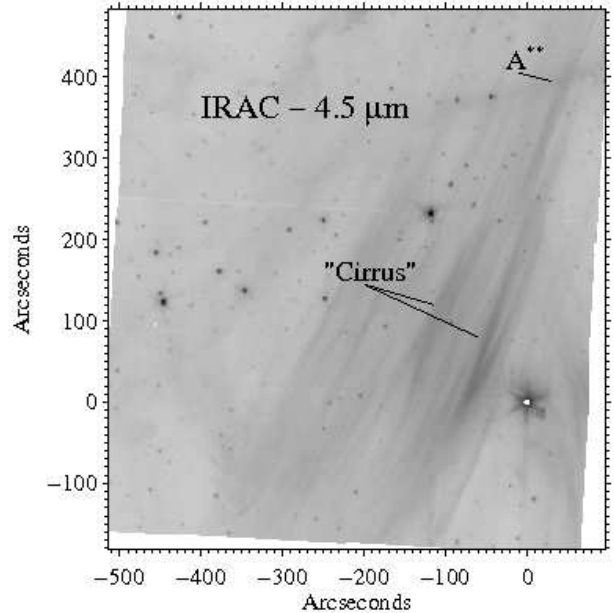}
\caption{\label{irac45} IRAC 4.5 $\mu$m image of the 
HD 34078 region with a logarithmic intensity scale to
enhance the nebular emission.  The ``cirrus'' features described
in \S3.1 are identified.  The 4.5 $\mu$m band shows these 
structures at a higher contrast than the shell structure 
seen in Figure 1.  
}
\end{center}
\end{figure}

\begin{deluxetable*}{ccccccc}
\tabletypesize{\small}
\tablecaption{Mid and far-IR photometry of bow shock region. \label{sptzbow}}
\tablewidth{0pt}
\tablehead{
\colhead{Position} & \colhead{RA (2000)\tablenotemark{a}}   &
\colhead{$\delta$ (2000)} & \colhead{F$_{8}$} & \colhead{F$_{24}$}  &
\colhead{F$_{70}$}  & \colhead{T$_{d}$\tablenotemark{b}}  \\ 
 &   \colhead{( $^{\mathrm h}$\, $^{\mathrm m}$\, $^{\mathrm s} )$} &
\colhead{(\arcdeg\, \arcmin\, \arcsec )}  & \colhead{(Jy)}   & 
\colhead{(Jy)}  & \colhead{(Jy)} & \colhead{(K)} \\
}
\startdata
S$_{E}$ & 05 16 20.98 & +34 18 52 & 0.508 $\pm$ 0.083 & 3.435  $\pm$ 0.229 &  1.945 $\pm$ 0.169 & 83.93 $\pm$ 2.06 \\
S$_{C}$ & 05 16 18.42 & +34 18 58 & 0.805 $\pm$ 0.103 & 4.286  $\pm$ 0.255 &  1.904 $\pm$ 0.167 & 88.76 $\pm$ 2.24 \\
S$_{W}$ & 05 16 16.83 & +34 18 48 & 0.675 $\pm$ 0.094 & 3.347  $\pm$ 0.225 &  1.805 $\pm$ 0.162 &  84.85 $\pm$ 2.16\\\\
 \enddata

\tablenotetext{a}{Shock pointings were defined from analysis of the 
24 $\mu$m MIPS data obtained near HD 34078.}
\tablenotetext{b}{Dust temperatures were derived from the ratio of 
24 to 70 $\mu$m flux density, F$_{24}$/F$_{70}$.  This temperature should be considered
a lower limit due to the saturation of the 24 $\mu$m map. }

\end{deluxetable*}

Using the mapping mode on the Infrared 
Spectrograph (IRS, Houck et al. 2004)\nocite{irs04}, we obtained
spectra clustered around previous $FUSE$ observations Pos3 (r16033024) and Pos4 (r16033536).  
IRS programs were carried out on 
2006 March 17 and 19 with a total exposure time of $\sim$~9600 seconds.
Data at all IRS wavelengths were acquired 
through both the 
high and low-resolution apertures (SL, SH, LL, and LH).  The spatial
coverage of the low-resolution aperture scan allows us to correlate
spectral variations with nebular morphology,
hence this work will focus on an analysis of the
low-resolution (Short-Low, 5~--~15$\mu$m; 
$R$~=~60~--~130) data.  Short-High and Long-Low spectroscopic
data were used as supporting observations.   
The post-BCD (pipeline version S13.2.0) reduced one-dimensional spectra were tagged according
to the central aperture position for registration with the associated
filaments observed in the imaging.  As the ``on-filament'' nebular
positions filled most of the Short-Low aperture, analyzing the full-slit extraction
at all pointings yielded information about both the spectral characteristics and the 
slit-filling fraction.

We also obtained mosaic maps of IC 405 in the three bands of the 
Multiband Imaging Photometer for $Spitzer$ (MIPS).  A description of the MIPS
instrument is given in~\citet{mips04}. The data (r16032768)
were obtained on 2006 February 28.
The post-BCD scan maps (S13.2.0) at 24 and 70 $\mu$m (Figures 2 and 3) were used to perform photometry of several 
nebular filaments 
and complement the
imaging at shorter wavelengths.    
The 24 and 70 $\mu$m maps are of high quality, but we note the 
inner 10~--~25\arcsec\ of the bow shock region near HD 34078
is likely saturated on the 24 $\mu$m Si:As array.
The 160 $\mu$m map contains gaps that
make measuring long-wavelength photometry very challenging, we will focus on 
the shorter wavelength MIPS channels here.   The 24 and 70 $\mu$m scans covers roughly
8\arcmin~$\times$~54\arcmin\ and 5\arcmin~$\times$~52\arcmin, respectively,  
both centered on $FUSE$ Pos4.

\subsection{$FUSE$}
IC 405 was observed by $FUSE$ on 2003 March 11~--~13 as part of the
D127 guest observer program~\citep{france04}.  Spectroscopy was 
obtained in the $FUSE$ far-UV bandpass (905~--~1187 \AA) through 
the low-resolution (30\arcsec~$\times$~30\arcsec; 
$\Delta\lambda\approx$ 0.33\AA) aperture (Moos et al. 2000).~\nocite{moos00}
The data acquired at positions Pos3 and Pos4 show H$_{2}$ emission 
lines and were used to guide the $Spitzer$ observations 
presented in this work.  We use the LiF2a (1087~--~1179\AA) 
channel to measure the strength of the strongest far-UV band of 
continuum pumped H$_{2}$ fluorescence, $\lambda\sim$~1100\AA, 
for comparison with the pure rotational emission lines in the IRS bandpass (\S4.2).
The $FUSE$ data were obtained in time-tagged mode
and reduced with the CalFUSE pipeline, version 2.2.3, with errors
determined from the associated count rate plots.  Both 
positions were observed for roughly 12.5 ksec.

\section{Analysis and Results: Imaging and Spectroscopy}
The 8.0 $\mu$m IRAC images showed the best combination of
S/N ($\sim$6~--~38 for the faintest to the brightest nebular features), 
spatial resolution ($\lesssim$2\arcsec; Fazio et al. 2004), 
and nebular structure with a field-of-view that
covered all of the subregions of interest.  A 
composite of the 8.0 $\mu$m images is shown in Figure~1.  These data provided
a reference frame for all of the image and spectroscopic analysis described below.  
The overall morphology of the 8.0 $\mu$m
images shows the brightest structures seen in the three-color optical 
image\footnote{T.A. Rector, http://antwrp.gsfc.nasa.gov/apod/ap011204.html; also 
presented in Figure 1 of France et al. (2004).}, displayed in Figure 1 for comparison with the IRAC data.
The ``shell'' structure seen to the East/Northeast of HD 34078 with a 
radius of $\sim$~400\arcsec\ observed in DSS images and in Plate IV of
\citet{herbig58} is observed at 8.0 $\mu$m.

\begin{deluxetable}{lcccc}
\tabletypesize{\small}
\tablecaption{8~$\mu$m IRAC photometry of nebular filaments. \label{sptz8fil}}
\tablewidth{0pt}
\tablehead{
\colhead{Position} & \colhead{Program}   & \colhead{RA (2000)}   &
\colhead{$\delta$ (2000)} & \colhead{Flux\tablenotemark{a}}  \\ 
 &   &   \colhead{( $^{\mathrm h}$\, $^{\mathrm m}$\, $^{\mathrm s} )$} &
\colhead{(\arcdeg\, \arcmin\, \arcsec )}     & \colhead{(Jy)}  \\ 
}
\startdata
A & r16033280           & 05 16 30.97 & +34 25 04 & 0.466 $\pm$ 0.079 \\
A$^{*}$ & r16033280 & 05 16 19.84 & +34 25 18 & 0.385 $\pm$ 0.072  \\
A$^{**}$ & r16033280           & 05 16 15.53 & +34 23 57 & 0.573 $\pm$ 0.088 \\
B & r16033792           & 05 16 16.88 & +34 32 21 & 0.543 $\pm$ 0.085 \\
\\
 \enddata


\tablenotetext{a}{The 8.0 $\mu$m images contain contributions from 
AEFs, H$_{2}$, and [\ion{Ar}{2}].  The 7.7 AEF contributes between
30~--~40\% of the 8.0 $\mu$m band flux in the region of peak filament 
brightness.}

\end{deluxetable}

Two bright nebular regions were chosen for analysis in this work, based on 
previous ultraviolet studies.
Filaments A and B were part of the~\citet{france04} far-UV study of IC 405.  
Data acquired at these positions showed a combination of fluorescent H$_{2}$ emission and
dust-scattered starlight. These observations provided 
detection of continuum pumped H$_{2}$ emission where the
rotational substructure was spectroscopically resolved at wavelengths shortward of Ly$\alpha$.
Filament A is broken up into three subregions: the 
main A knot with existing UV spectroscopy; $\sim$175\arcsec\ to the west, 
Filament A$^{*}$ is a clumpy feature located at the end of the 
Filament A structure; and Filament A$^{**}$ to the southwest was the brightest 
feature in the 70 $\mu$m data (these nebular features are labeled in Figure~1).  
The brightest features in the optical image are related to the regions
containing Filaments A and B at infrared wavelengths.  Filaments A and 
B are also the dominant features in the 6300~--~6750~\AA\ photographic
images presented in Herbig (1958; Plate III), where line emission from 
H$\alpha$, [\ion{N}{2}], and [\ion{S}{2}] is the primary source of the nebular 
brightness.\nocite{herbig58} 

The following subsections are divided into imaging and 
spectroscopic results.  The imaging is presented first, 
starting near HD 34078 and moving out into the northern filaments.
An analysis of the IRS spectroscopy follows, we determine the
relative contribution of the nebular species to the IRAC images and observe
position-dependent line strengths.
We note that the IRAC band overlap regions and spectroscopic programs were 
centered on Filaments A and B, thus six-band imaging and 
spectroscopy do not exist for all IC 405 subregions.    
A brief comparison between the imaging and spectroscopy of the different regions
is given in \S3.3.

\subsection{Imaging: IRAC and MIPS}

$Bow$ $Shock$ - The wake feature seen clearly in the 8.0 $\mu$m IRAC (Figure 4) and 
24 $\mu$m MIPS (Figure 5) images is thought to be enhanced dust emission 
produced by the interaction of HD 34078 with the ambient nebular medium~\citep{vanburen88}.
The bullet-tip interaction region is  seen most 
clearly in the 24 $\mu$m images (despite the saturation of the brightest
ridge), and three positions were chosen for analysis.  Moving from 
left to right on the image blow-ups, they are labeled Shock-East (S$_{E}$), 
Shock-Center (S$_{C}$), and Shock-West (S$_{W}$).  
The 8.0 $\mu$m data show an emission arc of diameter $\sim$~75\arcsec\
with an enhancement of factors of $>$~5 (in units of MJy sr$^{-1}$)
relative to the interfilament medium. 
The emission arc has a brightness peak 10~--~20\arcsec\ to the northeast 
of the stellar coordinates.  The 24 $\mu$m 
image shows a similar 75\arcsec\ arc, with higher contrast emission in the outer
edges of the arc.  The 24 $\mu$m data also suggest a  
connection between the shocked material and the cirrus features 
discussed below.  The 24 $\mu$m flux density transitions from the cometary arc into the cirrus features to the east of the star more smoothly than at 8 $\mu$m.
Both the 8 and 24 $\mu$m images show a clear spatial offset between the 
position of HD 34078 and the peak of the infrared emission.
The 4.5 $\mu$m  IRAC image is dominated by the saturated core of HD 34078 (Figure 6), 
but asymmetries in the PSF suggest that an enhancement exists
in the direction of stellar motion.   
An arc is detected in the 70 $\mu$m MIPS observations, which can be
seen faintly in Figure~3. 

\begin{figure}
\begin{center}
\hspace{+0.0in}
\includegraphics[angle=90,width=3.5in]{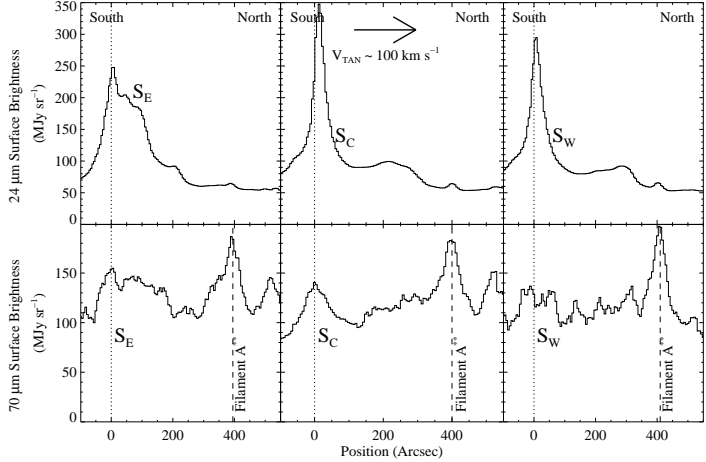}
\caption{\label{mipscuts} Spatial profiles of the 24 and 70 $\mu$m MIPS scans
of the southern region of IC 405.  The 24 $\mu$m data have been convolved to 
a resolution of $\approx$~18\arcsec\ for comparison with the 70 $\mu$m map.
The direction and magnitude of the stellar motion through IC 405 is indicated.
The 24 $\mu$m profiles outline the shock region while the 70 $\mu$m cuts show 
a weaker peak near the stellar position and a maximum near the Filament A$^{**}$
region.}
\end{center}
\end{figure}


\begin{deluxetable}{lcccc}
\tabletypesize{\small}
\tablecaption{24~$\mu$m MIPS photometry of nebular filaments. \label{sptz24fil}}
\tablewidth{0pt}
\tablehead{
\colhead{Position} & \colhead{Program}   & \colhead{RA (2000)}   &
\colhead{$\delta$ (2000)} & \colhead{Flux\tablenotemark{a}}  \\ 
 &   &   \colhead{( $^{\mathrm h}$\, $^{\mathrm m}$\, $^{\mathrm s} )$} &
\colhead{(\arcdeg\, \arcmin\, \arcsec )}     & \colhead{(Jy)}  \\ 
}
\startdata
A & r16032768           & 05 16 30.99 & +34 25 03 & 1.045 $\pm$ 0.126 \\
A$^{*}$ & r16032768 & 05 16 19.70 & +34 25 18 & 0.984 $\pm$ 0.123  \\
A$^{**}$ & r16032768           & 05 16 15.52 & +34 23 58 & 1.175 $\pm$ 0.134 \\
B & r16032768           & 05 16 18.17 & +34 32 27 & 0.892 $\pm$ 0.117 \\
\\
 \enddata


\tablenotetext{a}{IRS Long-Low spectra were analyzed to assess the emission line
contribution to the 24 $\mu$m band flux.  Discrete lines were found to contribute 
less than 8\% of emission in the 21.5~--~26.2 $\mu$m bandpass.}

\end{deluxetable}

We measured the variation in dust profile across the interaction region by 
taking spatial cuts through the 24 and 70 $\mu$m MIPS 
data, intersecting the S$_{E}$, S$_{C}$, and S$_{W}$ positions.
In order to make a fair comparison, the 24 $\mu$m data were 
convolved to a resolution of 18\arcsec\ and rebinned to the 
dimensions of the 70 $\mu$m scan.  The images were then 
aligned with a $\sim$~3\arcdeg\ rotation to account for 
spacecraft roll orientation and registered 
with their WCS data using a combination of image analysis tools (custom IDL
codes as well as ds9 from the Chandra X-Ray Center 
and the SSC data retrieval program Leopard). 
We found that the registration was accurate to roughly 2\arcsec\ in 
$x$ ($\sim$~RA) and 10\arcsec\ in $y$ ($\sim$~Dec).  
These profiles are shown in Figure 7 running from south to north
(left to right on the figure), intersecting the S$_{E}$, S$_{C}$, and 
S$_{W}$ coordinates listed in Table 1.  S$_{E}$, S$_{C}$, and 
S$_{W}$ are labeled in Figures 4 and 5, with dashed lines representing the
cuts in Figure 5.  
The 24 $\mu$m profiles 
show the dominance of the bow shock, again noting that the 
peak brightness regions (10~--~25\arcsec\ from HD 34078) are likely saturated.
The position of HD 34078 is indicated with the dotted line at 0 
and the direction of stellar motion through the nebula is labeled in the
center panel.
The S$_{E}$ profile shows the largest breadth due to
its association with the cirrus-like structures.
The 70 $\mu$m profiles show an enhancement at the shocked region, 
but rise to a maximum at the intersection of the nebular arm 
containing Filament A$^{**}$ ($y$~$\approx$~400\arcsec, 
marked with the dashed line in Figure 7).  The second peak in the 70 $\mu$m S$_{E}$
and S$_{C}$ profiles near 520\arcsec\ are associated
with the intersection of Filament A$^{*}$.  We also observe 
an additional peak 
at the intersection of the Filament B arm, but we do not show
a comparison as the image registration was of poorer quality.

\begin{figure*}
\begin{center}
\hspace{+0.0in}
\includegraphics[angle=90,width=6.5in]{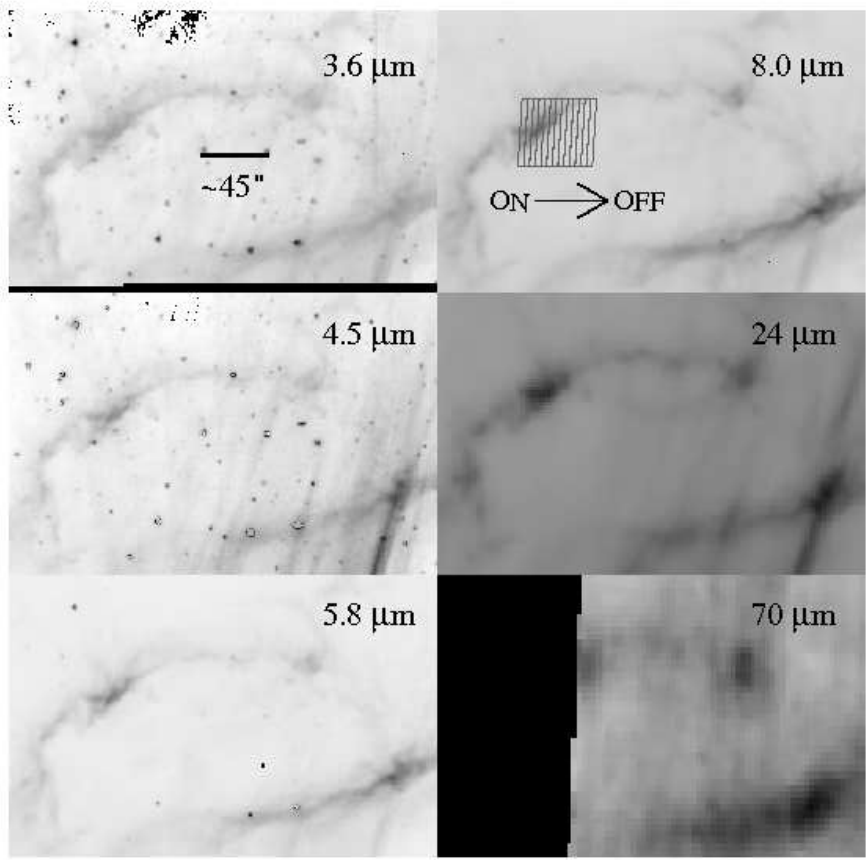}
\caption{
\label{6ircomp_01} Six-color imaging of the Filament A, A$^{*}$, and
A$^{**}$ regions.  Data from the four IRAC bands and two MIPS bands are shown
at their original resolution, with intensity scales intended to display the nebular structure.  
The black regions on the 3.6 and 70 $\mu$m 
images were not observed.  The overlays shown on the 8.0 $\mu$m image
represent the IRS-SL aperture scan direction and the ON/OFF convention used in this work.
This spectroscopy indicates that the AEFs make an 
important contribution to the nebular emission at positions of peak 
filament brightness. 
}
\end{center}
\end{figure*}

Flux ratio (F$_{24}$/F$_{70}$) profiles were made using these
extractions.  These profiles were interpolated onto 
a common spatial grid ($\approx$~4.7\arcsec/pixel) before comparison.  The registration 
in $y$ was the primary source of uncertainty for these comparisons.
These ratios were used to derive the dust temperature profile
in the shocked region, discussed in \S4.1.1.  Photometry 
of the S$_{E}$, S$_{C}$, and S$_{W}$ regions was performed on the  
8, 24, and 70 $\mu$m images following the procedure described below.  
The photometric values for the shock regions are listed in Table 1.

$The$ $HD$ $34078$ $Region$~--~The largest difference between the 8.0 $\mu$m and optical images
of IC 405 are the cirrus-like sheets of emission that are seen to the 
northeast of HD 34078 ($\Delta~x$~=~-250~--~0\arcsec, 
$\Delta~y$~=~0~--~300\arcsec\ in Figures 1, 4, 5, and 6).  This cirrus structure 
extends along filaments roughly from HD 34078 into the region containing
Filament A$^{**}$.  It may be possible to observe this substructure in the optical 
with high spatial resolution and a narrow field of view,  
imaging the nebula without strong contamination from the central star 
(i.e. - $HST$-ACS).

These cirrus sheets are the strongest features in the 4.5 $\mu$m IRAC
images of this region (Figure 6).  The region of patchy emission (-330\arcsec,-75\arcsec) 
seen at 8.0 $\mu$m is observed with a lower
relative strength at 4.5 $\mu$m.  MIPS 24 $\mu$m imaging of this region
finds similar relative intensities as those seen at 4.5 $\mu$m, strong 
cirrus sheets to the northeast of the star, with much lower relative
intensity features coincident with the patchy 8.0 $\mu$m 
and ``shell'' structure.  This
suggests that there is a common dust component responsible for the 
cirrus emission seen at 4.5, 8.0, and 24 $\mu$m, while there is an
additional component contributing to the 8.0 $\mu$m flux in the 
``shell'' region.  This additional 8.0 $\mu$m component is most likely
a combination of an increased contribution from the 7.7 and 8.6 $\mu$m
AEFs.  One of the results of the IRS scans of the bright filaments,
discussed below, is that the percentage contribution of AEFs to the 
total flux in a given imaging band increases in the 
brightest knots.  We  propose that the regions showing an enhanced
brightness in the 8.0 $\mu$m IRAC band relative 
to the dust emission at 4.5 and 24 $\mu$m are favorable for survival
of the AEF carriers.

$Filament$ $A$ - Filaments A$^{*}$ and A$^{**}$ were covered by all four IRAC imaging fields and the 
MIPS scans at both 24 and 70 $\mu$m.  A portion of Filament A was
lost at the edge of the 70 $\mu$m map, but covered in all other 
bands (Figure 8).  The 3.6 $\mu$m images of this region display a peak along the star-facing edge of the 
clouds containing the filaments.  The filaments are seen at 4.5 $\mu$m, 
though faint compared to the cirrus features seen near the star at 
this wavelength.  The 4.5 $\mu$m images were the weakest (in units
of MJy sr$^{-1}$) in our filamentary imaging.
We found the 4.5, 5.8, and 8.0 $\mu$m bands to be 
tracing the same spatial regions, with the 5.8 and 8.0 bands
enhanced relative to the 4.5 $\mu$m.  Again, this is attributed to additional 
flux from AEFs in the 5.8 and 8.0 $\mu$m bands.

The 24 and 70 $\mu$m MIPS images of Filament A show consistency with the 
8.0 $\mu$m reference image.  The 
24 $\mu$m map shows an additional flare structure pointing from Filament
A$^{**}$ to the northwest (RA~=~05$^{h}$16$^{m}$14.49$^{s}$,
$\delta$~=~+34$^{\circ}$24\arcmin47.0\arcsec, J2000).  
This feature is also seen weakly at 4.5 $\mu$m.  

\begin{figure*}
\begin{center}
\hspace{+0.0in}
\includegraphics[angle=90,width=6.0in]{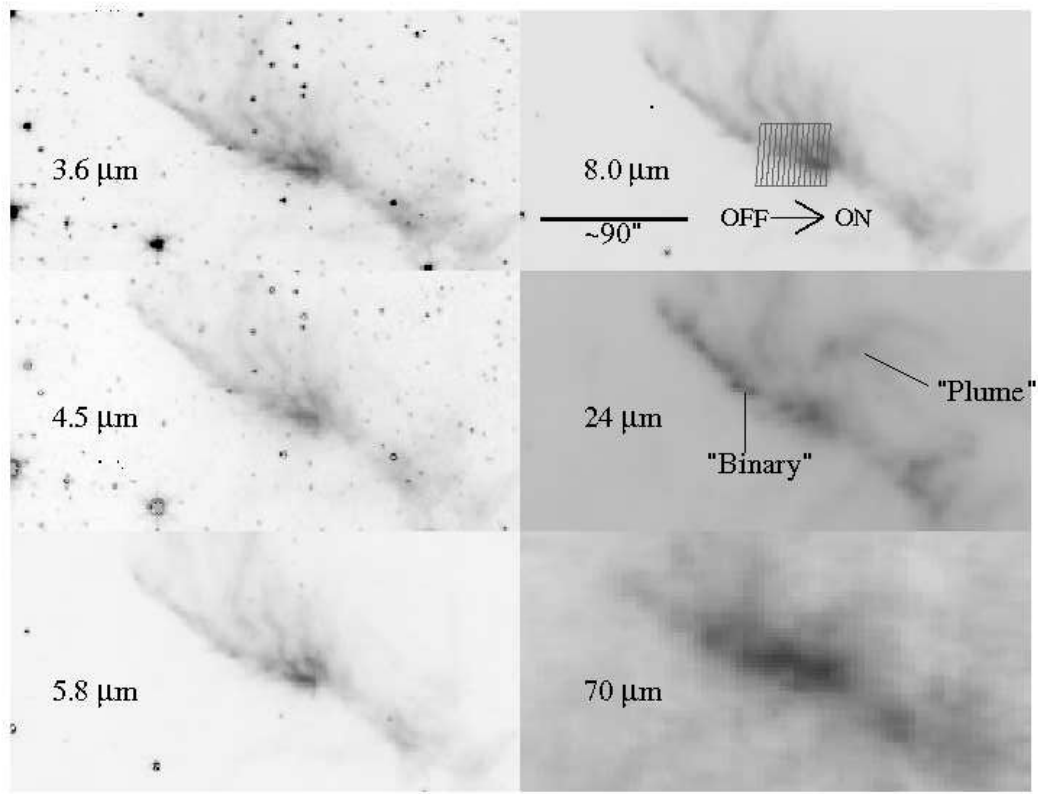}
\caption{
\label{6ircomp_02} Six-color imaging of the Filament B
region.  Data from the four IRAC bands and two MIPS bands are shown
at their original resolution. The images are qualitatively similar, 
but quantitative analysis shows that the South (HD 34078-facing) 
edge of filament is more extended at 24 $\mu$m, suggesting that
the carriers of the AEFs are being destroyed in the tenuous interfilament
medium.  The 24 $\mu$m map also shows a ``Plume'' to the 
north and a ``Binary'' feature to the northeast of the main Filament B emission, 
not seen as distinctly in other bands.  Again, the overlays shown on the 8.0 $\mu$m image
represent the IRS-SL aperture scan direction and the ON/OFF convention. 
}
\end{center}
\end{figure*}

Nebular photometry (at 8, 24, and 70 $\mu$m) of Filaments A and B
was performed and the flux densities are presented in Tables 2, 3, and 4.  
To ensure a fair comparison, the spatial resolution of the 
8.0 $\mu$m IRAC image and the 24 $\mu$m MIPS scan were degraded.
A two-dimensional Gaussian kernel was convolved with the data
to produce 8 and 24 $\mu$m images with an effective resolution of 
about 18\arcsec, comparable to that of the 70 $\mu$m map~\citep{mips04}.
The convolved images were integrated over a discrete
number of pixels and multiplied by the solid angle subtended by each pixel, 
3.4~$\times$~10$^{-11}$ sr pixel$^{-1}$ at 8 $\mu$m, 
1.5~$\times$~10$^{-10}$ sr pixel$^{-1}$ at 24 $\mu$m, and 
5.8~$\times$~10$^{-10}$ sr pixel$^{-1}$ at 70 $\mu$m.

We chose the photometric extraction regions based on the sizes of the 
filaments in the 8.0 $\mu$m data.  We found that the 
central cores of all four filament positions could be contained 
in a box of roughly 20\arcsec~$\times$~20\arcsec.  
A 20 pixel square 
used at 8 $\mu$m, a 10 pixel square was used in the 24 $\mu$m band, 
and a 5 pixel square extraction 	
was used for the 70 $\mu$m MIPS data.  
These extraction regions are of the same order as the projected size of the 
$FUSE$ LWRS aperture (30\arcsec~$\times$~30\arcsec) used at Filaments A and B (Figure 1).  
The photometric measurements were used to constrain the 
nebular dust temperature in IC 405, as discussed in \S4.1 and \S4.2.

\begin{deluxetable*}{lcccc}
\tabletypesize{\small}
\tablecaption{70~$\mu$m MIPS photometry of nebular filaments. \label{sptz70fil}}
\tablewidth{0pt}
\tablehead{
\colhead{Position} & \colhead{Program}   & \colhead{RA (2000)}   &
\colhead{$\delta$ (2000)} & \colhead{Flux}  \\ 
 &   &   \colhead{( $^{\mathrm h}$\, $^{\mathrm m}$\, $^{\mathrm s} )$} &
\colhead{(\arcdeg\, \arcmin\, \arcsec )}     & \colhead{(Jy)}  \\ 
}
\startdata
A & partially off scan map           & -- & -- &  -- \\
A$^{*}$ & r16032768 & 05 16 19.66 & +34 25 20 & 2.562 $\pm$ 0.193  \\
A$^{**}$ & r16032768           & 05 16 15.64 & +34 23 50 & 3.141 $\pm$ 0.214 \\
B & r16032768           & 05 16 17.74 & +34 32 32 & 2.356 $\pm$ 0.186 \\
\\
 \enddata



\end{deluxetable*}

$Filament$ $B$ - The Filament B region is located $\approx$~1000\arcsec\ north of HD 34078,
a part of what \citet{herbig58} labels the ``E nebulosity''.  This is
a position where $FUSE$ detected fluorescent emission from H$_{2}$ as
the most prominent spectral feature from Ly$\beta$ to 
1190~\AA~\citep{france04}.  Filament B was covered in all six imaging
bands used in this work as well as with spectroscopic
scans.  
Unlike the southern filaments described above, all four IRAC bands
appear to be tracing the same material in the Filament B region.
They show the same northeast-to-southwest elongated filamentary structure
with a peanut-shaped central concentration.  This peanut-shaped 
central structure, shown in all six IR imaging bands in Figure 9, 
is the center of Filament B (RA~=~05$^{h}$16$^{m}$16.88$^{s}$,
$\delta$~=~+34$^{\circ}$32\arcmin21.0\arcsec, J2000).  

The MIPS scans at 24 and 70 $\mu$m follow the structure seen in the 
IRAC bands, with a few noteworthy exceptions.  The MIPS data
show that the southern edge of the region occurs closer to the star at 24 $\mu$m
than at the IRAC wavelengths.  The 24 $\mu$m brightness increases 
gradually (B$_{24}$ 47~--~61 MJy sr$^{-1}$) over a $\sim$~30\arcsec\ 
transition region onto Filament B, while the 8.0 $\mu$m brightness 
rises more abruptly at the edge of the filament, increasing in brightness 
by more than a factor of two (B$_{8}$ 18~--~40 MJy sr$^{-1}$) over 
a 15\arcsec\ interface region.  Destruction of small grains and the
carrier of the 7.7 $\mu$m AEF in the more tenuous environment 
south of Filament B could produce this effect.  We may be seeing the 
erosion of this cloud in progress, the strong UV field of HD 34078 
photoevaporating the carbonaceous grains and weakly shielded molecules.
This idea could be tested by searching for the remnants of these species 
([\ion{C}{2}]~$\lambda$158 $\mu$m for example; van Dishoeck 2004)
with future far-IR spectroscopic instrumentation, 
such as $Herschel$-PACS~\citep{pacs04}.  

The 24 $\mu$m MIPS data also reveal a plume of material to the 
north-northwest of the main knot of Filament B
(RA~=~05$^{h}$16$^{m}$16.74$^{s}$,
$\delta$~=~+34$^{\circ}$33\arcmin09.5\arcsec, J2000).  The IRAC images show
hints of emission at this location, but only at 24 $\mu$m is the contrast
high enough to identify this plume as a separate feature.  It is worth
noting that this plume seems to be ``in the shadow'' 
of Filament B, analogous to the position of the flare filament 
seen in the shadow of Filament A$^{**}$ at 24 $\mu$m.  The other
notable feature seen in the MIPS imaging is the binary structure seen to the
northeast of Filament B at 24 $\mu$m
(RA~=~05$^{h}$16$^{m}$21.75$^{s}$,
$\delta$~=~+34$^{\circ}$32\arcmin41.9\arcsec, J2000).  
The two components of this feature are not spatially resolved at 70 $\mu$m.
The ``Plume'' and ``Binary'' features are labeled on Figure 8.

Integrated flux densities for Filament B were measured from the 
IRAC 8.0, MIPS 24, and MIPS 70 $\mu$m images.
The values are listed in Tables 2, 3, and 4.  We find the 
24/70 $\mu$m ratio to be essentially identical in Filaments A$^{*}$, A$^{**}$, and B 
(0.379~$\pm$~0.005), indicative of similar dust temperatures in the three regions (\S4.2).  

\begin{figure}
\begin{center}
\hspace{+0.0in}
\includegraphics[angle=90,width=3.75in]{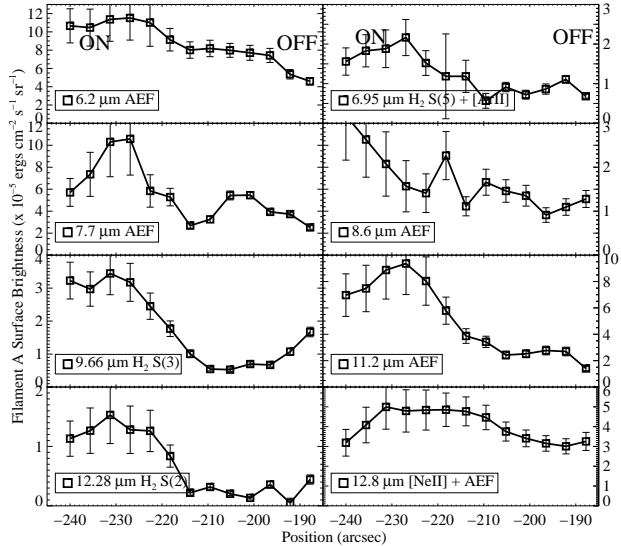}
\vspace{+0.15in}
\caption{
\label{p4irs_01} Filament A line strengths measured with the Short-Low
module on the IRS, as a function of position.  The coordinates 
correspond to the 8.0 $\mu$m reference image, Figure 1.
The map goes from 
East (-245\arcsec, on the core of the filament) to West 
($>$ -220\arcsec, off of the peak of the 8.0 $\mu$m filament emission).
}
\end{center}
\end{figure}

\subsection{Spectroscopy: IRS}

$Filament$ $A$ - The IRS scan maps of Filament A have been used
to measure the variations in emission characteristics from dense 
knots to tenuous nebular material.  
The spatial and spectral coverage of the Short-Low (5~--~15 $\mu$m) 
module is best suited to studying the change in 
AEF signature while scanning across the region of interest in full-slitwidth
steps.  Filament A was mapped from east to west, 
with the eastern portion of the map being located on the brightest parts
of Filament A, while the western portion was off the brightest Filament A
emission (Figure 10).  

\begin{figure}
\begin{center}
\hspace{+0.0in}
\includegraphics[angle=90,width=3.5in]{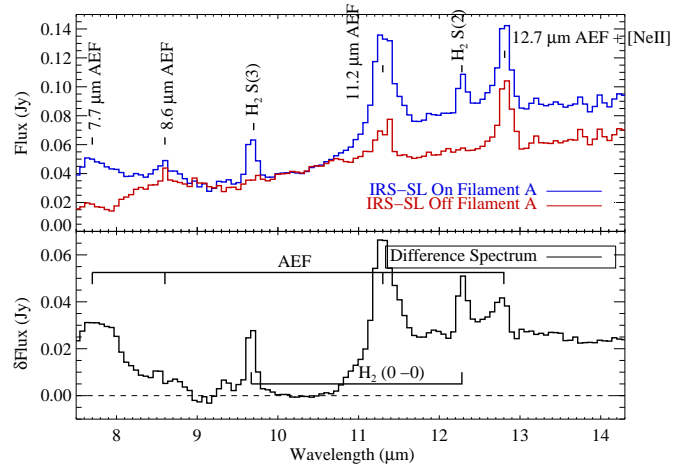}
\caption{
\label{p3irs}  IRS spectra of the ON-filament and
OFF-filament regions near Filament A.  The OFF-Filament 
was scaled up to the ON-filament flux level in the 10 $\mu$m region. 
The  ON-filament (top panel, blue) spectrum shows the 7.7 and 11.2 AEFs, 
H$_{2}$, and the 11~--~14 $\mu$m ``plateau'' to be enhanced compared to the 
OFF-filament (top panel, red) pointing.  
The bottom  panel shows the difference of the ON and OFF-filament 
spectra to illustrate the differences.}
\end{center}
\end{figure}

All observed emission lines follow the same generic trend:  They are brightest near the 
peak of the 8.0 $\mu$m reference image, and their intensity
falls off with distance from the peak (Figure 10).
The AEFs at 6.2, 8.6, and 11.2 were detected clearly at all Filament A
pointings.  Band substructure in the 6.2 $\mu$m feature is discussed in \S4.4.
Line strengths were measured by integrating the spectra over the
wavelengths of interest (in the case of the 7.7 $\mu$m band, the
1$^{st}$ and 2$^{nd}$ orders were joined before measurement).  The continuum contribution was removed 
by subtracting 
a linear fit to the continuum over the wavelengths spanned by the line.
The 7.7 $\mu$m AEF was seen to be the most variable spectral feature, 
ranging from marginally detectable off the filament to a wide ($\Delta\lambda~\sim$~1~$\mu$m) band 
at the peak of the filament brightness.  
For all IRS-SL spectral lines, we conservatively assumed a 20\% 
flux error on each measurement (line and continuum) which resulted to total line 
strength uncertainties of a few to approximately 30\%.

Figure 11 shows examples of the low-dispersion spectra acquired at Filament A, 
showing several aromatic and H$_{2}$ emission lines.
The spectra obtained on the peak of the 8 $\mu$m emission also 
show a broad ``plateau'' feature from 11~--~14 $\mu$m that is not
seen in the spectra obtained off the filament.  
This plateau was first seen by $IRAS$~\citep{cohen85}, and $Spitzer$
detected a similar 10.5~--~14.5 $\mu$m 
feature in NGC 7023, however in that nebula the individual features wash out into
a broad $\Delta\lambda$~$\approx$~2 $\mu$m feature farther from the 
central star~\citep{werner04b}.
A similar effect with distance from the illuminating star was seen by 
\citet{cesarsky00} in the reflection nebula Ced 201.  
In the case of IC 405 however, the entire spectroscopic scan of Filament A is
roughly equidistant from HD 34078, thus we are most likely not observing an evolution
with radiation field.  
The imaging indicates that the spectra sample less dense regions as the scan moves off
the filament.   This result would be consistent with the 11~--~14 $\mu$m plateau 
being controlled by density\footnote{We note that while IR flux density 
images are representative of the underlying 
density structure, the IRAC flux density does not strictly reflect the 
nebular density~\citep{heitsch06}.}.  

\begin{figure}
\begin{center}
\hspace{+0.0in}
\includegraphics[angle=90,width=3.75in]{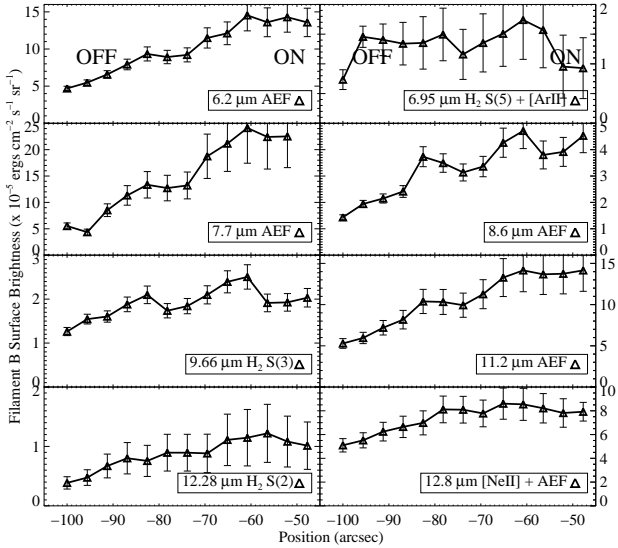}
\vspace{+0.15in}
\caption{
\label{p4irs_02} Filament B line strengths measured with the Short-Low
module on the IRS, as a function of position.  The coordinates 
correspond to the 8.0 $\mu$m reference image, Figure 1.  The map goes from 
East (-105\arcsec, off of the filament) to West 
($>$ -70\arcsec, near the peak of the 8.0 $\mu$m filament emission).
}
\end{center}
\end{figure}

Mid-IR spectra of the Orion Bar~\citep{bregman89} indicate that
the molecules responsible for the 11~--~13 $\mu$m plateau are independent of
the carriers of the narrower AEFs. 
In IC 405, the AEFs, as well as the lines of 
H$_{2}$ and 12.8 $\mu$m AEF+[\ion{Ne}{2}] (described below), show an 
increase in peak flux with the plateau strength.  Figure~11
illustrates the changes in these features.  The top panel shows a comparison 
between spectra observed at the peak of the filament brightness
and spectra observed off the filament.  The ``off'' spectrum is scaled
to the 10.0~--~10.5 $\mu$m  region of the ``on'' spectrum.
The difference of the two is plotted in the lower panel.  The 
relative strengths of the 7.7 $\mu$m feature, the H$_{2}$ lines, and the 
11.2 and 12.8 $\mu$m features are all enhanced with the 11~--~14 $\mu$m 
plateau.  

Several atomic emission lines produced in ionized gas were 
observed in the Short-High observations clustered on the brightest region
of Filament A, but we only report on the species that appear in 
more than 15 of the 25 pointings.  These spectra show strong [\ion{Ne}{2}] $\lambda$12.81 $\mu$m 
at all positions.  In the low dispersion spectra of Filament A, the 
average central wavelength of the 12.8 $\mu$m feature is observed at 
$\langle\lambda_{obs}\rangle$~=~12.817~$\pm$~0.005 $\mu$m, consistent with the 
observed emission coming from ionized neon as opposed to the 12.7 $\mu$m AEF.  
In addition to [\ion{Ne}{2}], 
[\ion{S}{3}] $\lambda$18.71 $\mu$m is detected in most of the high dispersion data.
We note that it is interesting to find these 
energetic ions (ionization potentials are $>$ 21 eV for \ion{Ne}{2} and \ion{S}{3}) spatially aligned with the emission
lines of H$_{2}$, described below.  One might naively expect that 
the molecules would be destroyed in a region with highly excited ions, yet they
are observed within the same 4.7\arcsec~$\times$~11.3\arcsec\
(IRS Short-High aperture) region.  This may suggest that
the H$_{2}$ in IC 405 resides in dense globular reservoirs, as in planetary
nebulae~\citep{huggins02,speck03,lupu06,hora06}.  Narrow band 
near-IR imaging in the rovibrational emission lines of H$_{2}$ ($\lambda$~$\sim$~2 $\mu$m)
would be useful in testing this suggestion.  Finally, we detect emission lines
at 12.94 and 13.14 $\mu$m in the majority of the Short-High spectra at a signal to noise ratio
comparable to the H$_{2}$ lines discussed below. 
The 12.94 $\mu$m line remains unidentified, but we tentatively identify the 13.14 $\mu$m feature
as the \ion{O}{1} $\lambda\lambda$13.139~--~13.157~$\mu$m multiplet.  
Hot pixels in the Short-High module do not appear to be the source of the unidentified
lines, however, the 13.14 $\mu$m line does appear as a single pixel in some of the pointings.
Given the possibility of hot pixel contamination and the fact that 
the excitation process for the neutral oxygen line is unclear, 
we do not feel that this is a conclusive detection.

The pure rotational emission lines of H$_{2}$ are seen at the 
on-filament pointings.  The (0~--~0) S(5), S(3), and S(2) lines are 
strongest in the core of Filament A, becoming weaker 
in all directions outside of the brightest 20\arcsec. 
The S(4) line was only weakly observed in one pointing.  
The S(5) line of H$_{2}$ is likely blended with [\ion{Ar}{2}], as we measure a line center that
lies between the rest wavelengths for the molecular and atomic lines,
$\lambda_{obs}$~=~6.946~$\pm$~0.010, where as $\lambda_{S(5)}$~=~6.91 and 
$\lambda_{[ArII]}$~=~6.99 $\mu$m.  The continuum subtraction for the
S(3) and S(2) lines is somewhat uncertain due to the variability of
the AEFs on whose wings they reside.  Due to concerns about what 
effects this may have on the determination of the 
H$_{2}$ rotational temperature (T(H$_{2}$)), we also used observations
of the brightest filamentary regions with the Short-High IRS module
to make clean measurements of the H$_{2}$ (0~--~0) S(2) $\lambda$12.28 
and S(1) $\lambda$17.03 $\mu$m lines when they are 
seen at high S/N.  The determination of T(H$_{2}$) is discussed in \S4.2.

The contributions from the 6.2, 7.7, and 8.6 $\mu$m AEFs to the 
5.8 and 8.0 $\mu$m IRAC bands  were measured in Filament A.  
Line strengths were determined as described above, and compared to the total spectroscopic flux measured 
in the IRS data over the 5.8 and 8.0 $\mu$m IRAC imaging windows~\citep{irac04,werner04}.
We found that the 6.2 $\mu$m feature was responsible for $\sim$ 33~--~57\%
of the emission in the 5.8 $\mu$m IRAC band, with a peak in the AEF 
contribution coincident with the peak in 5.8 $\mu$m brightness.
The relative contributions from H$_{2}$ S(5)+[\ion{Ar}{2}] 
($\lambda$~$\approx$~6.95~$\mu$m), and the  
7.7 and 8.6 $\mu$m AEFs were measured in the IRAC 8.0 $\mu$m bandpass.
The  H$_{2}$ S(5)+[\ion{Ar}{2}] was negligible off the filament and 
rose to a maximum of 3\% at the peak 8.0 $\mu$m brightness.  The 
8.6 $\mu$m AEF contributes 1.0~--~1.5 \% at all positions, while the
7.7 $\mu$m AEF showed a strong increase on the 
filament.  At the western end of the IRS scan (the ''OFF'' filament
position) the 7.7 $\mu$m feature contributes about 8\% of the 
flux in the 8.0 $\mu$m  image, increasing to 34\% at the peak of the 8.0 $\mu$m 
brightness.  Following this argument and noting the brightness
of the 3.6 $\mu$m images relative to 4.5 $\mu$m, we consider it 
likely that a substantial fraction of the 3.6 $\mu$m brightness 
is due to emission from the 3.3 $\mu$m AEF, although we note that
the scattered starlight component increases at shorter wavelengths as well.

\begin{figure}
\begin{center}
\hspace{+0.0in}
\includegraphics[angle=90,width=3.5in]{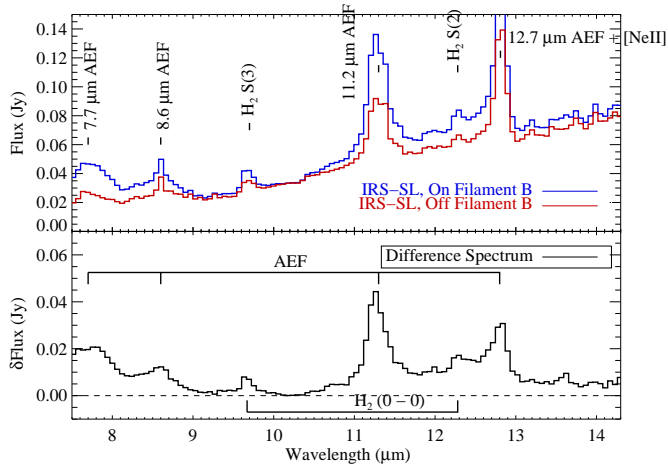}
\caption{
\label{p4irs_03} IRS spectra of Filament B. The blue spectrum in the
top panel was acquired on the peak of the 8.0 $\mu$m emission, while the red
spectrum was acquired off of the brightest region of the filament.  
The OFF-filament spectrum was scaled to the ON-filament flux between 10~--~10.5 $\mu$m.
The difference of the two spectra is plotted in the lower panel.  The strengths of the pure rotational H$_{2}$
lines and the 11~--~14 $\mu$m plateau are observed to be less variable than at Filament A (Figure 11).  
}
\end{center}
\end{figure}

$Filament$ $B$ - The dust, H$_{2}$, and atomic lines observed in the
Short-Low spectra of Filament B follow two behaviors.  The 6.2, 7.7, 8.6, and
11.2 AEFs show a roughly linear increase in brightness as the scan 
moves onto the filament, while H$_{2}$ S(5)+[\ion{Ar}{2}], H$_{2}$
S(3) and S(2), and AEF+[\ion{Ne}{2}] $\lambda$12.8 $\mu$m show a rise
in strength then flatten out across the filament.  This is shown in 
Figure 12.  Examples of the spectra obtained at Filament B are shown in Figure 13.  
The 11~--~14 $\mu$m plateau is also observed at the filament position, 
although with less variability than at Filament A. 
The Short-High spectra were again used to cleanly measure the
S(2) and S(1) lines near the peak of the Filament B brightness.
As in Filament A, we detect [\ion{Ne}{2}]~$\lambda$12.81 and [\ion{S}{3}] $\lambda$18.71 $\mu$m at the 
majority of the pointings.  The $\lambda$12.94 and $\lambda$13.14 $\mu$m features are observed as well.

Following the procedure outlined for Filament A, we measured the 
contribution of line emission to the 5.8 and 8.0 $\mu$m images.  The 
6.2 $\mu$m AEF is an appreciable fraction of the 5.8 $\mu$m IRAC band 
brightness, ranging from 40\% at the eastern end of the scan map
(OFF filament) to 46\% at the peak of the Filament B.  The
H$_{2}$ S(5)+[\ion{Ar}{2}] blend makes up 2.8\% of the 
nebular flux off the filament, dropping to 1.5\% on the 
center of the filament.  This is opposite the behavior seen at Filament A, 
where the relative H$_{2}$+[\ion{Ar}{2}] contribution increased 
with filament brightness.  The 7.7 and 8.6 $\mu$m AEFs 
relative contribution to the 8.0 $\mu$m image increases substantially 
with position.  Both lines contribute roughly 
2.5 times more to the ON filament positions than they do in the more
diffuse nebular pointings.  The 7.7 $\mu$m AEF rises from 13 to 35\%  
while the 8.6 $\mu$m feature increases from 2 to 5\%.

\begin{deluxetable*}{l|lcc}
\tabletypesize{\small}
\tablecaption{Filament A and B molecular line strength and dust brightnesses. \label{sptzlines_vert}}
\tablewidth{0pt}
\tablehead{
 & \colhead{$Spitzer$}  & \colhead{Filament A}   & \colhead{Filament B} \\
 & \colhead{$FUSE$}  & \colhead{Pos3}   & \colhead{Pos4} 
}
\startdata
Species & & & \\
H$_{2}$ & $\langle$(0~--~0) S(3) Brightness$\rangle$\tablenotemark{a} & 2.84 $\pm$ 0.62 & 2.14 $\pm$ 0.25 \\
 & $\langle$(0~--~0) S(2) Brightness$\rangle$ & 1.22 $\pm$ 0.23 & 1.08 $\pm$ 0.12 \\
 & UV  Brightness\tablenotemark{b} & 2.00 $\pm$ 0.60 &  0.40 $\pm$ 0.16 \\
 & T(H$_{2}$) (K)\tablenotemark{c} & 418$^{+108}_{-73}$  & 387$^{+46}_{-39}$ \\
  &    &   \\
AEFs & $\langle$6.2 $\mu$m Brightness$\rangle$ & 8.73 $\pm$ 2.23 & 10.13 $\pm$ 3.39 \\
 & $\langle$7.7 $\mu$m Brightness$\rangle$ & 8.52 $\pm$ 2.31 & 9.39 $\pm$ 3.89 \\
 & $\langle$8.6 $\mu$m Brightness$\rangle$ & 1.69 $\pm$ 0.66 & 3.30 $\pm$ 1.03 \\
 & $\langle$11.2 $\mu$m Brightness$\rangle$ & 5.05 $\pm$ 2.79 & 10.58 $\pm$ 3.16 \\
  &    &   \\
Blends & $\langle$H$_{2}$ + [\ion{Ar}{2}] 6.95 $\mu$m Brightness$\rangle$ & 1.23 $\pm$ 0.50 & 1.29 $\pm$ 0.28 \\
 & $\langle$AEF + [\ion{Ne}{2}] 12.8 $\mu$m Brightness$\rangle$ & 4.04 $\pm$ 0.77 & 7.35 $\pm$ 1.15 \\
 &    &   \\
Dust & F$_{24}$ (Jy) & 1.045 $\pm$ 0.126 & 0.892 $\pm$ 0.117 \\
 & F$_{70}$ (Jy) & -- & 2.356 $\pm$ 0.186  \\
 \enddata
\tablenotetext{a}{H$_{2}$ brightnesses are averages over six ON filament
pointings at each postion. AEF and Blend brightnesses are averages over all IRS-SL pointings
at each filament.  They are in units of (x 10$^{-5}$ ergs cm$^{-2}$ s$^{-1}$ sr$^{-1}$).}
\tablenotetext{b}{Measured over the $\sim$ 1100~\AA\ line complex (1097.7~--~1108.0~\AA).}
\tablenotetext{c}{Measured from the ($I(S(3))/I(S(2))$) ratio. 
Filament A shows evidence for temperature fluctuations to T(H$_{2}$)~$\sim$~700 K (see \S4.2).}
\end{deluxetable*}

\subsection{Regional Comparison}

We can compare the general characteristics of Filament B to those
of Filament A (Table 5).  The H$_{2}$ rotational emission lines
are brighter in Filament A, 
which could indicate a higher H$_{2}$ density in Filament A or
possibly a higher photoelectric grain heating rate 
due to the relative proximity of HD 34078~\citep{pdrs,abel05}.
Similarly, we observe 24 $\mu$m emission, 
presumably emission from grains, to be brighter in Filament A.
Again, we would expect the grain population nearer the illuminating
star to display an enhanced brightness.   
However, the AEFs show the opposite behavior, they are seen 
to be anti-correlated with the strength of the far-UV radiation field.  
The 6.2, 7.7, 8.6, and 11.2 $\mu$m AEFs all have a
position averaged brightness higher in Filament B.  The 8.0 $\mu$m 
photometric measurements are higher in Filament B, consistent 
with a considerable contribution from a brighter 7.7 $\mu$m feature.  
As the AEF carriers and small ($a~\lesssim$~30 \AA) grains are thought to be 
heated stochastically by the absorption of a 
single UV photon~\citep{draine01}, one might expect the AEF flux to be higher 
near HD 34078. We suggest that
the opposite behavior may be caused by increased photodestruction of the 
AEF carriers in the more intense UV radiation field at Filament A.  
The large column density of hydrogen, relative to the aromatics, 
may place the H$_{2}$ in a self-shielding regime 
while the AEF carriers are more easily dissociated by the UV radiation field of HD 34078.	
An analogous situation is seen in UV illuminated PDRs where 
H$_{2}$ is observed while the lower column density, unshielded, CO is destroyed~\citep{luhman96,luhman97}.

\begin{figure}
\begin{center}
\hspace{+0.0in}
\includegraphics[angle=90,width=3.25in]{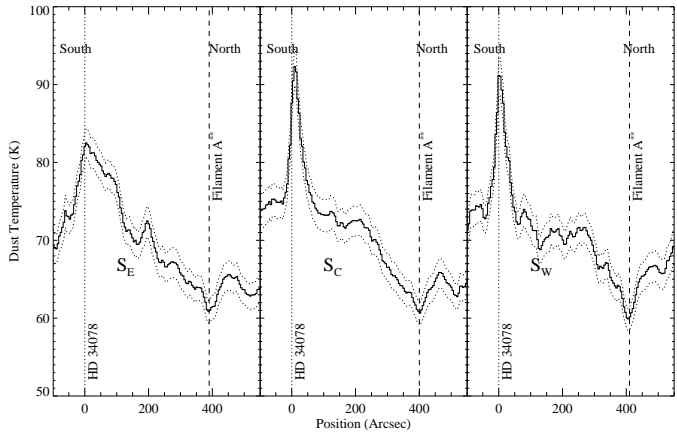}
\caption{\label{tempcuts} Dust temperature profiles of the southern portion of IC 405, from 
HD 34078 (at the origin) to Filament A$^{**}$.  The error ranges are plotted as dotted lines. 
The temperature profiles are created by taking
the ratio of spatial extractions from the 24 and 70 $\mu$m MIPS scans.  The dust temperature
is seen to rise sharply at the location of the bow shock, 10~--~25\arcsec\ north of HD 34078.
The minimum dust temperature of $\sim$~60 K is observed at the position of Filament A$^{**}$.  This
temperature is characteristic of the dense filaments in IC 405.}
\end{center}
\end{figure}

IRS mapping observations of the shock region were not
part of the P20434 program, but predictions about the spectral 
characteristics of the region can be made.  H$_{2}$ (0~--~0) S(3) $\lambda$9.66 $\mu$m, 
(1~--~0) S(1) $\lambda$2.12 $\mu$m, and [\ion{O}{1}] $\lambda$63 $\mu$m should 
be readily observable in this region~\citep{vanburen90}.  \citet{draine83}
make predictions about cooling through additional rotational lines of H$_{2}$, as
well as CO, OH, H$_{2}$O, and [\ion{C}{1}]. 

\section{Discussion}

\subsection{Cometary Bow Shock Associated with HD 34078}
Far-IR bow shocks are found around roughly 30\% of galactic O and B 
runaway stars~\citep{vanburen95,noriega97}.  These stars are
ejected from the their formation site by interactions, primarily with either a supernova 
binary companion or members of the cluster in which they were formed~\citep{huthoff02}.  
When one of these high space velocity ($\geq$~30 km s$^{-1}$) stars
encounters even moderately dense interstellar material, a bow shock is formed
by ram pressure confinement of the OB stellar wind as it moves through the ambient medium~\citep{vanburen88,noriega97}.
UV radiation from the star heats the shocked gas and dust, 
producing excess far-IR emission in these shock regions, which has been observed by $IRAS$.  

HD 34078 was a bow shock candidate in the $IRAS$ survey of~\citet{vanburen95}, showing an 
excess brightness at 60 $\mu$m, but no shock morphology could be detected.  A higher
spatial resolution (1\arcmin) follow up to this survey~\citep{noriega97} did not resolve
the shock.  It is only the imaging capabilities of $Spitzer$ 
($\leq$~6\arcsec\ at $\lambda~\leq$~24 $\mu$m) 
that allow us to unambiguously identify the HD 34078 bow shock. 
Figures 4 and 5 show the stellar region at 8 and 24 $\mu$m, identifying the
three shock regions of interest, as described in \S3.1.  Inspection of the 
images makes clear the ``cometary'' morphology~\citep{vanburen90}, though the 
$\sim$~75\arcsec\ diameter arc is more extended than traditional ultracompact \ion{H}{2}
regions that are interpreted as bow shocks~\citep{maclow91}.

\subsubsection{Temperature Structure}
The dust in the bow shock region should be heated by the UV radiation field of HD 34078, and we
use spatial extractions of the 24 and 70 $\mu$m MIPS maps to confirm this prediction.  The
F$_{24}$/F$_{70}$ ratio cuts have been converted  into dust temperature profiles as a function 
of position following the modified black body prescription described in \S4.2.  
The dust temperature
profiles for the S$_{E}$, S$_{C}$, and S$_{W}$ regions are shown in Figure 14.  These cuts extend from roughly
100\arcsec\ south of HD 34078 to 550\arcsec\ north of the star, intersecting the 
Filament A$^{**}$ region.  We observe the $\sim$~60 K filament temperature derived below for Filament A$^{**}$, and 
see the rise in dust temperature near HD 34078, increasing sharply in the inner 100\arcsec\ to a 
peak that is coincident with the location of the cometary morphology on the images.  The peak dust 
temperatures for the shock regions are T$_{d,peak}^{S_{E}}$~=~82.5$\pm$1.9, 
T$_{d,peak}^{S_{C}}$~=~92.3$\pm$2.4, and T$_{d,peak}^{S_{W}}$~=~91.2$\pm$2.5, respectively.
These peak temperatures are lower limits as the 24 $\mu$m MIPS data were saturated along 
a portion of the bright bow shock ridge.  The dust temperatures derived from the photometric 
measurements (given in Table 1) of the shock region average over the peak 24 $\mu$m brightness and the surrounding emission. 

\begin{figure}
\begin{center}
\hspace{+0.0in}
\includegraphics[angle=90,width=3.5in]{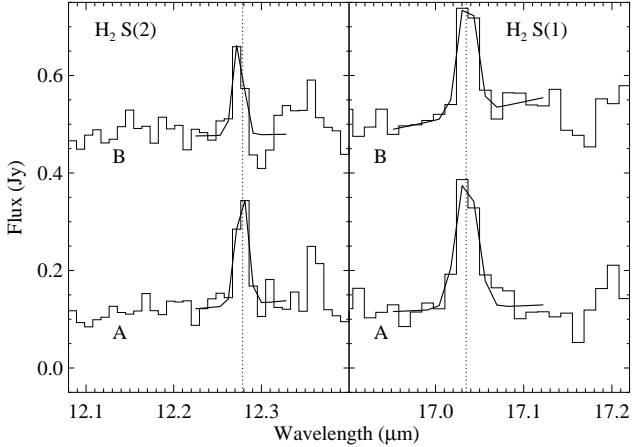}
\caption{\label{irsshp3p4comp} Short-High spectra of the H$_{2}$ (0~--~0) 
S(2) and S(1) pure rotational emission lines at 12.28 and 17.03 $\mu$m.  
An offset of 0.4 Jy was added to the Filament B spectra for this figure.
Ratios of the measured line strengths ($I(S(2))/I(S(1))$) are used to 
derive the H$_{2}$ rotational temperature, T(H$_{2}$).
The $I(S(2))/I(S(1))$ temperature in Filament B agrees with that
derived from Short-Low measurement of the $I(S(3))/I(S(2))$ temperature
(T$^{B}$(H$_{2}$)~$\approx$~400 K).  However, the Short-High 
observations of Filament A show evidence for non-uniform excitation corresponding
to a temperature of roughly 700 K. 
}
\end{center}
\end{figure}

The dust temperatures we derive are factors of 2~--~3 higher
than seen in the original $IRAS$ bow-shock survey~\citep{vanburen88}.  
The high temperature of the HD 34078 shock can be related to the high velocity of the star
and the high spatial resolution of these observations.
The $\sim$100 km s$^{-1}$ velocity of HD 34078 is outside the 
range of velocities considered by~\citet{vanburen95}.  HD 34078 has
the largest one-dimensional proper motion (within the errors) of these shock 
candidates ($pm_{y}$ in Table 2 of van Buren et al. 1995).  
The high spatial resolution provided by $Spitzer$-MIPS allows 
us to spatially resolve the hottest shock regions.  The larger $IRAS$ beam (5\arcmin\ for
van Buren et al 1995; and 1\arcmin\ for Noriega-Crespo et al. 1997) sampled 
the entire shock region, diluting the warmer inner regions with cooler dust
on the exterior of the shock.  Averaged over 5\arcmin\ 
around HD 34078 (-90~--~210\arcsec\ of Figure 14), the 
temperature derived for S$_{C}$ decreases to 76 K.  

\subsubsection{Bow Shock Density: H$_{2}$ Excitation and UV Dust Extinction} 
The morphological and thermal evidence places the HD 34078/IC 405 interface
in the cometary bow shock family, and we can use the analytic models of 
van Buren \& McCray (1988) and van Buren et al. (1990) to explore other topics related to the 
IC 405 system.~\nocite{vanburen88,vanburen90}  
An estimate of the nebular density enables us to calculate the column density and 
UV optical depth associated with the bow shock and explore the implications for existing studies.

The distance from the star to the shock region can be calculated from Equation (2) 
of (van Buren et al. 1990; see also van Buren \& McCray 1988), assuming an 
existing knowledge of the nebular density.  Alternatively, 
we can measure the star-shock distance directly from the IRAC and MIPS observations,
and solve for the density. Let $l$ be the measured 
distance from the star to the shock structure in the direction of motion, 
$\approx$15\arcsec\ on the 8.0 $\mu$m IRAC image.  At the assumed distance of 450 pc, this corresponds 
to a separation of 1.77~$\times$~10$^{15}$ cm.  Rewriting equation (2) of~\citet{vanburen90},  
the density ($n_{H}$~=~2$n_{H_{2}}$~+~$n_{HI}$~+~$n_{HII}$) in the region of IC 405 
through which HD 34078 is currently moving is given by
\begin{equation}
n_{H}~=~(3.03~\times~10^{38})~\dot{m}_{*,-6}~v_{w,8}~\mu_{H}^{-1}~v_{*,6}^{-2}~l^{-2}~~cm^{-3}
\end{equation}
where $\dot{m}$$_{*,-6}$ is the stellar wind mass loss rate in units of 
$\times$~10$^{-6}$~M$_{\odot}$~yr$^{-1}$,
$v_{w,8}$ is the terminal velocity of the stellar wind in units of $\times$~10$^{8}$ cm s$^{-1}$, 
$v_{*,6}$ is the stellar velocity in units of $\times$~10$^{6}$ cm s$^{-1}$,  and $\mu_{H}$
is the dimensionless mass per H-nucleus.  We take $\mu_{H}$ to be 1 and assume 
a stellar velocity of $\sim$~100 km s$^{-1}$~\citep{blaauw54,boisse05}.  
The stellar mass loss rate (log($\dot{m}$$_{*}$)~=~-9.5) and terminal wind velocity 
(800 km s$^{-1}$) are taken from~\citet{martins05}.  
Following this procedure, we calculate a nebular density of $n_{H}$~=~244 cm$^{-3}$.

Rewriting Equation (7) of~\citet{vanburen90}, the column density of the swept-up material is given by
\begin{equation}
N~=~(1.95~\times~10^{19})~\dot{m}_{*,-6}^{1/2}~v_{w,8}^{1/2}~\mu_{H}^{-1/2}~v_{*,6}^{-1}~n_{H}^{1/2}~~cm^{-2}
\end{equation}
with the parameters defined as before.  This calculation yields a column density
of material piled-up by the intrusion of HD 34078, $N$~=~4.8~$\times$~10$^{17}$~cm$^{-2}$.
This value is very similar to the H$_{2}$ column density found by~\citet{boisse05}
for the 
hot absorption component along the HD 34078 line of sight.
They find $N(H_{2},J \geq 5)$~=~3.3~$\times$~10$^{17}$ cm$^{-2}$, comparable to their
model prediction of 2.4~$\times$~10$^{17}$ cm$^{-2}$.  Our derived column density 
of the bow shock region supports the hypothesis that the high-lying states of 
H$_{2}$ observed on the HD 34078 sightline are produced in a hot layer near the star, 
swept-up by the passage of HD 34078, and excited by stellar UV photons. 

The sightline to HD 34078 is seen to be highly reddened compared
with other regions of IC 405~\citep{france04}.  This is inferred from the blue rise of the
ratio of nebular surface brightness to stellar flux (S/F$_{*}$) at far-UV
wavelengths (900~--~1400 \AA).  The $Spitzer$ data allow us to determine if this
enhanced reddening can be attributed to the material associated with the shock region.
Equation (6) of~\citet{vanburen88} gives the UV optical depth ($\tau_{UV}$)
of the swept-up dust.  Assuming that the majority of the dust survives the interaction
with HD 34078 and that the swept-up material has a velocity of a few km s$^{-1}$, 
$\tau_{UV}$ is of the order 0.01.  This relatively small value of $\tau_{UV}$ is 
insufficient to explain the observed S/F$_{*}$ in IC 405.  In other words, the 
dust responsible for the differential extinction towards IC 405 is not associated
with the bow shock.  It is more likely that this dust is located in the 
translucent cloud along the HD 34078 sightline.
This cloud is observed to produce the majority of the cold ($J$~=~0, 1; T(H$_{2}$)~=~77 K) 
H$_{2}$ absorption~\citep{boisse05}, and we conclude that this cloud is also responsible
for the majority of the extinction towards HD 34078.

\begin{figure}[b]
\begin{center}
\hspace{+0.0in}
\vspace{+0.4in}
\includegraphics[width=3.25in]{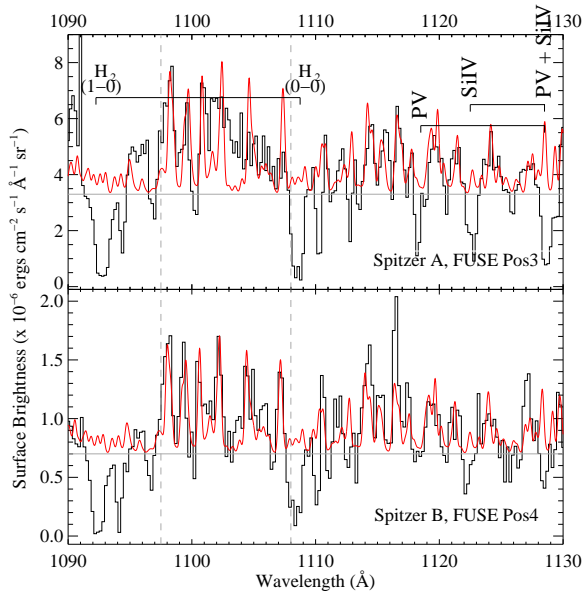}
\caption{\label{fusep3p4comp}\vspace{+0.0in} $FUSE$ observations of Filaments A 
(above) and B (below).  H$_{2}$ fluorescence is observed at each position 
imposed on the dust scattered continuum of HD 34078, 
providing direct evidence for a population of photo-excited nebular gas. 
A synthetic spectrum of H$_{2}$ emission, 
scaled to the flux level of the data, is overplotted.  
Lyman band H$_{2}$ (labeled 0~--~0 and 1~--~0) and photospheric absorption lines  are also observed.  
}
\end{center}
\end{figure}

\subsection{H$_{2}$ and Dust Temperatures in Filaments A and B}
Observationally, little attention has been paid to IC 405 
in comparison with other bright PDRs (NGC 2023, NGC 7023, IC 63, and the Orion Bar).  
Here we establish the dust and H$_{2}$ rotational temperatures of the outer filaments.

$H_{2}$ -- Rotational temperatures can be measured from 
H$_{2}$ emission line ratios in the IRS bandpass~\citep{rosenthal00,allers05}.
This temperature directly probes the bulk of the 
molecular mass in PDRs and has been a widely used diagnostic since
a large number of mid-IR observations were made available by 
$ISO$ (see references in the review by van Dishoeck 2004 and \S1.1).~\nocite{thi99,habart04}
We have used the IRS Short-Low measurements of the position averaged H$_{2}$
S(3) and S(2) emission line strengths and Short-High observations of the central 
cores of Filaments A and B to determine T(H$_{2}$)
for these regions.   Only the brightest six Short-Low pointings corresponding
to the ON position in Figures 10 and 12 were considered.  
The use of the Short-High
module allows us to separate the (0~--~0) S(2) $\lambda$12.28 $\mu$m
and S(1) $\lambda$17.03 $\mu$m lines from nearby spectral features.
The S(2) and S(1) line strengths were measured from the two cleanest
(defined as a combination of line S/N and a low, featureless background)
Short-High observations at each filament.  The S(2) and S(1) spectra from Filaments A and B 
are shown in Figure 15.

T(H$_{2}$) was determined by combining the measured line ratios,
$I(S(2))/I(S(1))$ and $I(S(3))/I(S(2))$, with energy levels and transition 
probabilities from the literature.  This temperature determination assumes that the pure rotational lines are optically thin and thermally populated, with an equilibrium ortho-to-para H$_{2}$ ratio of 3.  Internal and foreground 
attenuation is also assumed to be negligible, 
which is supported by the finding that the reddening towards the nebula is 
essentially zero at far-UV wavelengths~\citep{france04}.

The energy levels from~\citet{mandy93} give $\Delta E_{21}/k$~=~667 K,   
and the Einstein-$A$ radiative transition probabilities are 
$A_{S(2)}$~=~2.76~$\times$~10$^{-9}$ s$^{-1}$ and 
$A_{S(1)}$~=~4.76~$\times$~10$^{-10}$ s$^{-1}$~\citep{avals98}.  The ratio 
of level degeneracies, $g_{S(2)}/g_{S(1)}$, is 9$/$21.   
Using these parameters, we obtain rotational temperatures of 
T$^{A}_{21}$(H$_{2}$)~=~692~$\pm$~55 K and T$^{B}_{21}$(H$_{2}$)~=~381~$\pm$~28 K
for Filaments A and B, respectively.
These temperatures are higher than one would expect for a cold 
interstellar cloud, but are similar to recent PDR observations that find 
typical H$_{2}$ rotational temperatures
in the range of a few hundred degrees K~\citep{thi99,habart04,allers05}.

Noting the added uncertainties in the continuum levels discussed above, 
we also calculated the H$_{2}$ rotational temperature using position
averaged strengths of the S(3) and S(2) lines observed in the low-dispersion
spectra. Including the values for $A_{S(3)}$, $\Delta E_{32}/k$, from the references 
given above with $g_{S(3)}$, we found T$^{A}_{32}$(H$_{2}$)~=~418$^{+108}_{-73}$~K and 
T$^{B}_{32}$(H$_{2}$)~=~387$^{+46}_{-39}$~K.  The agreement at Filament B
is remarkable (381 vs. 387), but the Filament A temperatures differ 
by almost 300 K.  

We interpret variations seen at Filament A as differing relative populations in adjacent rotational states, 
indicative of the influence of non-thermal excitation by far-UV photons.
The distribution of excited robvibrational states produced in the fluorescent cascade 
differs from a thermal distribution, and this is reflected in the observed 
emission line ratios~\citep{black87}.
If the filament structure is clumpy, the narrower field-of-view of the Short-High
aperture (as opposed to the position averaged Short-Low data used to determine
T$_{32}$(H$_{2}$)) could be dominated by a knot with a larger non-thermal emission fraction.
Two tests of the Short-Low data have been performed to explore this possibility.
We have used the individual H$_{2}$ S(3) and S(2) line strengths to look at
T$_{32}$ as a function of position.  This shows that the derived H$_{2}$ rotational
temperature displays excursions to greater than 700 K, albeit at the lower signal to noise positions
not considered in the average line strengths used to determine T$_{32}$ above.  
Secondly, we looked at temperature variations along the Short-Low slit at a pointing that 
spatially overlapped the Short-High spectra used.  We measured a higher temperature 
(T$^{A}_{32}$(H$_{2}$)~$\gtrsim$~520 K) along the star-facing edge of Filament A, 
coincident with the Short-High positions, suggesting that the line ratios used to determine 
the H$_{2}$ rotational temperature may be altered by UV photons in this region.


The majority of rovibrational emission lines that would indicate 
fluorescent pumping fall at wavelengths shorter than the $Spitzer$-IRS 
bandpass, thus IRS spectroscopy alone may not be sufficient 
to constrain the excitation mechanism of nebular H$_{2}$.
In IC 405 however, direct evidence for non-thermal (fluorescent) excitation of H$_{2}$ has
been presented by~\citet{france04}.  $FUSE$ observations
clearly revealed the far-UV spectrum of fluorescent molecular hydrogen at 
Filaments A and B.   These spectra are shown in Figure 16.
Filament A is shown in the top panel, where the emission  
from H$_{2}$ is seen imposed upon dust-scattered light from 
HD 34078.  We note that the fluorescent lines at Pos3 display 
a broadening in excess of the instrumental width, as has been seen in the 
H$_{2}$ emission spectra of other 
PDRs~\citep{france05a}. 
HD 34078 is clearly identified as the excitation source by  
the highly excited photospheric features of \ion{P}{5} and \ion{Si}{4}
seen in the scattered light.
As has been shown for fluorescent H$_{2}$ emission near the 
Trapezium~\citep{france05b}, a star with such photospheric ions~\citep{pellerin02} 
is the most likely to have the requisite far-UV flux ($\lambda$~$\lesssim$~1110~\AA) to 
produce observable UV fluorescence.  The $FUSE$ spectrum of Filament B
is shown in the bottom panel, overplotted with a synthetic H$_{2}$ 
spectrum~\citep{france04}.   It is interesting
to note that the synthetic spectrum was created with a rotational temperature of 400 K, 
only attempting to find qualitative agreement with the data.
The fluorescent H$_{2}$ was integrated over
the strongest far-UV line complex (1097.7~--~1108.0~\AA) after subtracting the 
contribution from dust scattered light.  Integrated surface brightnesses are  
quoted in Table 5. 

Fluorescent excitation by HD 34078 can also be observed in 
the highly excited H$_{2}$ absorption lines on the HD 34078 sightline~\citep{boisse05}.
As non-thermal emission clearly plays a role in IC 405, we propose that the 
average rotational temperatures of the filaments is $\sim$~400 K, while  
UV-pumping has produced non-uniform excitation.  The result of this
excitation has been to alter the observed line ratios at the Short-High
pointings used in Filament A, leading to the observed temperature deviation. 


$Dust$~--~Thermal dust temperature (T$_{d}$) can be determined
from mid and far-IR photometry, made widely available
by $IRAS$~\citep{sodroski87,schnee05}.  Measurements of T$_{d}$
fit two or more IR data points, assuming emission from the dust cloud is described
by a modified black-body curve.  The black-body function 
is typically modified by a power-law emissivity function that
is representative of the grain composition.

We have used the 24 and 70 $\mu$m MIPS observations of IC 405 
to derive the dust temperature in the three filaments where the
scans overlap (A$^{*}$, A$^{**}$, and B), as well as across the 
bow shock region (\S4.1.1).  
The contribution of discrete spectral features to the photometric
flux in these bands has been considered.  The weak 
spectral features in IRS Long-Low (20~--~30~$\mu$m) 
observations contribute less than $\sim$~8\%
of the 24 $\mu$m band flux at all pointings.  The contribution
of line emission to the 70 $\mu$m brightness is more difficult to 
constrain without the aid of spectroscopy.  We note that [\ion{O}{1}]
$\lambda$63~$\mu$m lies at the edge of the MIPS 70 $\mu$m filter. 
[\ion{O}{1}] may be strong in IC 405~\citep{vand04}, but
without direct measurement of the nebular SED at these wavelengths, we do not
attempt to correct for emission lines.

The observed F$_{24}$/F$_{70}$ photometric 
ratio, computed from Tables 3 and 4, was fit using 
a modified black body spectrum, with a dust 
emissivity proportional to $\nu^{2}$. The choice for the 
dust emissivity law was determined by the theoretical 
dust absorption opacity wavelength dependence 2.92$\times$10$^{5}$ 
($\lambda$/$\mu$m)$^{-2}$, valid in the 20~$\mu$m$\leq \lambda \leq$700~$\mu$m 
range for the average diffuse ISM~\citep{li01}.  
The spectral slope in the 24 and 70 $\mu$m filter 
bandpasses does not make a significant contribution to the resulting brightness ratios, 
the 
correction being significantly smaller than our measurement errors for T$_{d}$~$\geq$~40 K.


The F$_{24}$/F$_{70}$ ratios were nearly identical in the three
filaments, 0.379 (\S3.1), and the inferred dust temperatures reflect this.  
The mean temperature for the nebular dust in IC 405 is 63 K.
Individually, the observed filament dust temperatures are 
T$_{d}^{A^{*}}$~=~62.8~$\pm$~1.5, T$_{d}^{A^{**}}$~=~62.6~$\pm$~1.4, and 
T$_{d}^{B}$~=~62.7~$\pm$~1.6 K.
These grain temperatures are roughly a factor of 3.5 higher than
those found for molecular clouds in Perseus, Ophiucus, and Serpens
($\langle$T$_{d}$$\rangle$~=~17 K; Schnee et al. 2005), suggesting that
the grains have processed the UV photons from HD 34078.~\nocite{schnee05}
Our derived temperature for the IC 405 filaments is similar 
to the lower limit of the dust temperature (55 K) in the Orion Bar PDR~\citep{lis98}. 
What may be surprising about the IC 405
dust measurements is the similarity of the temperature at 
Filaments A$^{*}$ and A$^{**}$ (roughly equidistant from HD 34078) with that
of Filament B.  Single photon heating effects should be important for grains with
$\lesssim$ 1000 carbon atoms in the intense radiation field of HD 34078~\citep{draine01}.
This may be the cause of the constant temperature distribution in the nebular filaments.

\begin{figure}
\begin{center}
\hspace{+0.0in}
\includegraphics[angle=90,width=3.5in]{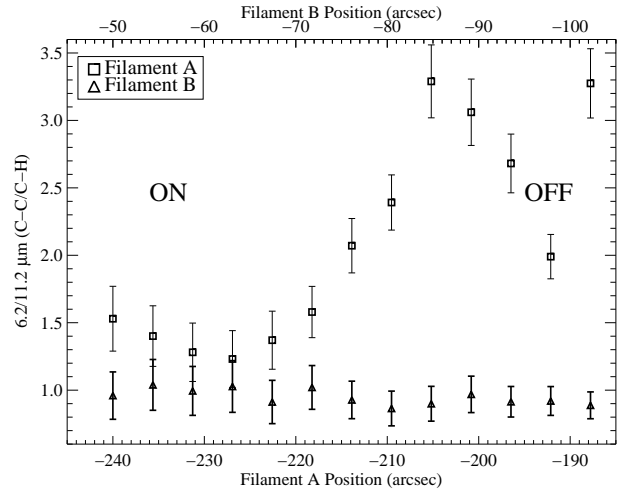}
\caption{
\label{rat_62_112} IRS measurements of the ratio of 6.2 and 
11.2 $\mu$m line strengths in Filaments A and B.  
Their ratio is an 
indication of the ionization state of the aromatics in these regions.  The relative ionization 
fraction increases off the core of Filament A, but is seen to be constant
at Filament B. $Note:$ The Filament B data is displayed ON (-50\arcsec) to OFF (-105\arcsec) 
for comparison with Filament A in this plot.
}
\end{center}
\end{figure}

\subsection{Ionization and Molecular Environment}
There is general consensus that the AEFs are produced by a superposition of 
vibrational modes of large carbonaceous molecules~\citep{verstraete01,peeters02}.  
Laboratory and theoretical studies have shown that
the typical interstellar aromatic emission spectrum is composed of a combination
of neutral and ionized carriers with a distribution of sizes
(see the review by van Dishoeck, 2004, and references therein).
These studies have shown that cations (positively charged ions) 
have enhanced 6~--~10 $\mu$m features arising from stronger C~--~C
stretching modes.  Neutral molecules have stronger C~--~H bands, 
producing the majority of the aromatic emission at 3.3 
and 10~--~14 $\mu$m~\citep{hony01,bregman05}.  
The ratio of 6.2/11.2 $\mu$m emission bands can be indicators
of ionization~\citep{uchida00,bakes01} and molecular structure~\citep{chan01,hony01}.  

IRS scans of the two filaments in IC 405
allow us to study the change in ionization state with environment.
Assuming that Filaments A and B are located at roughly the same heliocentric distance
as HD 34078, we can constrain the relative radiation fields at the two positions.
Variations in the spectral properties along Filaments A and B
can be correlated with change in filament structure/density $and$ changing radiation field.  
Previous studies have primarily focused on the variations of AEFs 
with radiation field strength~\citep{verstraete96,peeters02,werner04b}.
Our results for molecular ionization fraction as a function of position are 
shown in Figure 17.  The scan moves off the filament to the right, 
the position coordinates have been calculated to correspond to the
images presented in this work. 
The 6.2/11.2 ratio is displayed as squares for Filament A and triangles 
for Filament B.  We observed an increase in ionization fraction 
as the scan moves from the central knot to the diffuse
nebula at Filament A, going from roughly 1.4 to 3.3.  
This can be understood as an increase in the contribution
from ionized carriers in less dense regions at a constant
radiation field~\citep{bakes01}.

\begin{figure}
\begin{center}
\hspace{+0.0in}
\vspace{+0.05in}
\includegraphics[width=2.75in,angle=90]{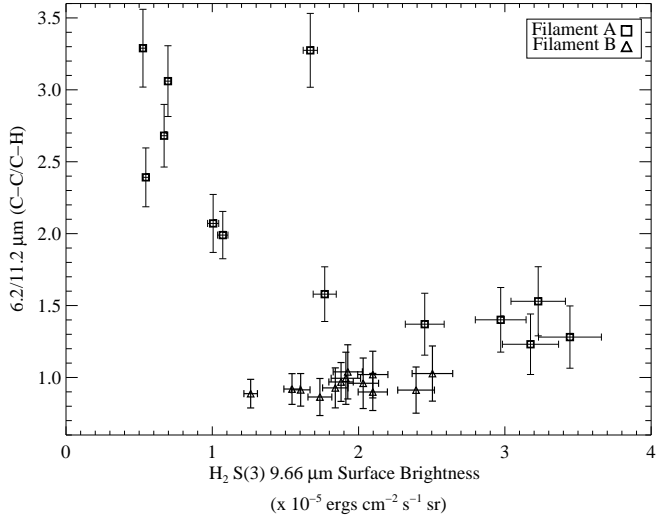}
\caption{
\label{h2pahrat} The ionization measure of the AEF carriers (6.2/11.2 $\mu$m)
as a function of strength of the molecular hydrogen emission at 9.66 $\mu$m.
The correlation observed at Filament A suggests that ionization of the aromatics
increases in lower density environments at given radiation field~\citep{bakes01}.
No trend is observed at Filament B.}
\end{center}
\end{figure}

The ionization fraction is seen to be constant with position in Filament B. 
The baseline 6.2/11.2 $\mu$m ratio is lower in Filament B than in 
Filament A ($\sim$ 0.9 vs 1.4), suggesting that the overall
ionization fraction is smaller in Filament B.  Filament B
is roughly twice the angular distance from HD 34078
and consequently receives about a quarter of the UV flux present at 
Filament A, consistent with the $FUSE$ observations presented in \S4.2.  
While the ionization fraction declines with diminishing UV flux, 
there might not be a one-to-one relationship.
This conclusion is derived in connection to the assertion made in 
\S3.3 that Filament A shows indirect evidence for enhanced AEF carrier
destruction relative to Filament B.  Using the Filament B region as a baseline, 
our observations are consistent with the photodestruction of the aromatic molecules
responsible for the AEFs being a multi-step process where dissociation is preceded
by ionization.   Additional observations of PDRs are available in the
$Spitzer$ and $ISO$ archives, and we will explore similarities with IC 405
in future work.

The AEF ionization measure can also be 
correlated with the environment in which H$_{2}$ emission is produced.  
A comparison of the H$_{2}$ emission strength and AEF ionization is shown
in Figure 18.  The S(3) rotational line was used as representative of 
H$_{2}$ emission because of its brightness and separation from 
other lines.  We observe the AEF ionization fraction to increase
as the molecular hydrogen line strength decreases in Filament A.  
Assuming that H$_{2}$ is representative of density, 
these results also indicate that the AEF carrier population is
more ionized in more tenuous regions at a given UV radiation field~\citep{bakes01}.
Combining this result with the spatial dependence of the molecular
ionization described above confirm that in strongly illuminated PDRs, 
the 6.2/11.2 $\mu$m ratio is a good diagnostic of the local ionization conditions.
This result is consistent with the \citet{bakes01} prediction for regions
with a ratio of ionizing radiation field to electron density ($G_{o}$/$n_{e}$) of
$\sim$~10$^{4}$ cm$^{3}$.  No correlation is observed at Filament B.

\subsection{Substructure in the 6.2 $\mu$m Feature -- Evidence for PANHs?}
In addition to providing information about the ionization state,
the 6.2 $\mu$m feature can be a diagnostic of the AEF carrier structure.
\citet{peeters02} have performed a survey of the 6.2 $\mu$m feature
in a variety of astronomical objects, and have determined empirical 
subclasses for this band.  Lines with peak wavelengths
near 6.22 $\mu$m form Class A, while lines peaking between 6.24 and 6.3
$\mu$m make up Classes B and C.  Most \ion{H}{2} regions, 
reflection nebulae, and non-isolated Herbig Ae/Be stars belong to 
Class A, while isolated Herbig Ae/Be stars and most planetary 
nebulae belong to Class B.  They find red-peaked ($\lambda_{o}$ $\sim$~6.3 $\mu$m) lines to be composed
of ``pure'' PAHs, while blue-peaked ($\lambda_{o}$ $\sim$~6.2 $\mu$m) lines have N-atoms
substituted at inner lattice sites.  

\begin{figure}
\begin{center}
\hspace{+0.0in}
\includegraphics[angle=90,scale=.7]{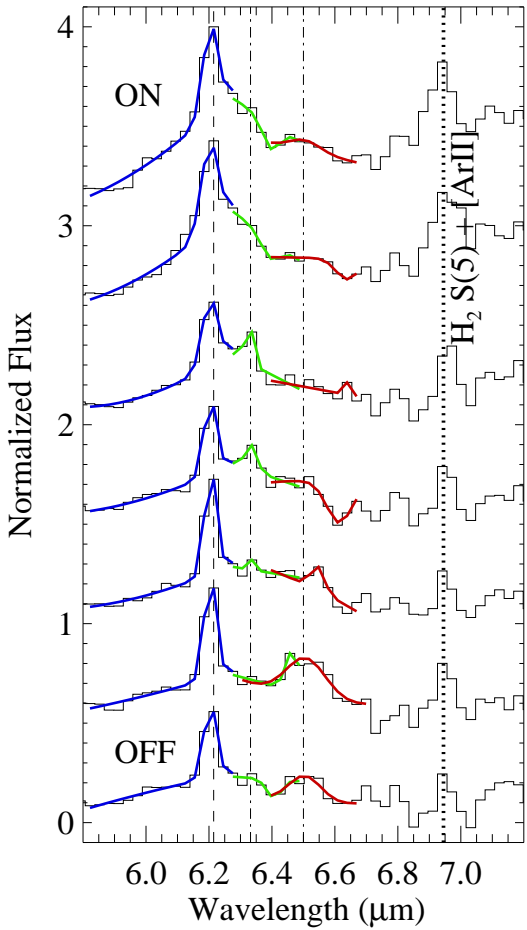}
\caption{
\label{redcomp62} IRS Short-Low observations of the 
6.2 $\mu$m AEF in Filament A.  The spectra have been normalized to the peak of
the on-filament 6.2 $\mu$m band and offset by factors of 0.5 for this display.
The data show a variation in the 6.2 $\mu$m profile from the core of the filament (ON, top) to the 
edges (OFF, bottom).  The spectra maintain a Class A 6.2 $\mu$m 
peak~\citep{peeters02}, but develop red substructure away from the
core of the filament.  Three Gaussians are fitted to the spectra showing the
main peak (blue), the second peak seen at 6.33 $\mu$m (green), and the 
broad 6.5 $\mu$m ``hump'' (red).  The 6.5 $\mu$m Gaussian on the
second OFF spectrum (offset~=~+0.5) was fit over a slightly expanded wavelength
range to show the feature more clearly.
}
\end{center}
\end{figure}

\citet{hudgins05} carry the nitrogenation hypothesis further, 
determining the necessary site for N-substitution in order to reproduce the 
observed 6.2 $\mu$m wavelength and ruling out increasing molecular size
 alone as the explanation for the observed line profiles.
They find that nitrogen substitution is required to
reproduced the observed profiles, labeling this PAH subclass
polycyclic aromatic nitrogen heterocycles (PANHs).  They conclude that
PAHs or PANHs with N-substitution on the edges of the molecule 
produce the emission at $\lambda$~$\geq$~6.3 $\mu$m, while PANHs with 
the N-atom substituted in the interior of the molecule are responsible for 
the Class A spectra observed by \citet{peeters02}.

With this observational and theoretical evidence in mind, we
report a tentative detection of band substructure to the 
red of the main 6.2 $\mu$m peak at Filament A.  Filament A
spectra are shown in Figure 19 from 5.8~--~7.2 $\mu$m.
The spectra are
dominated by a main Class A feature at $\lambda_{max}$~=~6.210~$\pm$~0.007
$\mu$m .  At the peak filament brightness position, 
we observe an enhanced red wing to the main feature.  This wing develops a
distinct second peak at 6.33 $\mu$m as the overall 6.2 band decreases in 
strength away from the central core of the filament.  An additional
red ``hump'' appears further to the red ($\lambda~\sim$~6.50 $\mu$m)
becoming stronger with increasing distance from the peak filament brightness position.
Three Gaussians are shown overplotted on the data in Figure 18 
representing each component.  The H$_{2}$+[\ion{Ar}{2}] blend is 
seen near 6.95 $\mu$m.  

The red features are qualitatively similar to the  
coronene cation (C$_{24}$H$_{12}^{+}$)
spectra shown in Figure 3 of \citet{hudgins05}.  
The substructure is most pronounced in the same region where 
the ionization fraction is seen to increase.  One interpretation is that 
processing by UV photons is altering the PAH population in the interface 
between the dense filament and the more tenuous nebular regions.
We only see red substructure in one Filament B spectrum.  
Filament B is generally observed to have an 
unmodified Class A 6.2 $\mu$m band, $\lambda_{max}$~=~6.219~
$\pm$~0.004 $\mu$m.  We note that a 6.35 $\mu$m feature
was reported as a red shoulder of the 6.2 $\mu$m band in $ISO$
observations of M17--SW~\citep{verstraete96}, interpreted as 
large PAHs under the influence of a strong UV radiation field.
However, \citet{peeters02} report no detection of a 6.35 $\mu$m feature.


\section{Summary}

We have presented new $Spitzer$ observation of IC 405, 
an emission/reflection nebula in Auriga illuminated by the 
runaway O9.5 star HD 34078.
We have used IRS spectroscopy to assess the morphological differences seen in 
four band (3.6, 4.5, 5.8, and 8.0 $\mu$m) imaging obtained with IRAC.
The combination of the IRAC data with 24 and 70 $\mu$m MIPS maps
of the nebula reveal wavelength dependent morphology in several
subregions.  We attribute these differences to variations
in the relative strengths of AEFs and emission from grains.  
In addition to the AEFs, the spectroscopy revealed emission from the
pure rotational lines of H$_{2}$ and atomic fine structure lines.
The emission bands/lines were observed to vary 
in both absolute and relative strength as a function of position
at two bright nebular filaments.

Dust temperatures in three nebular filaments were derived from 24 and 70 $\mu$m
photometry.  The dust temperature was a constant 63 K, independent of the distance
from the illuminating star.  
A similar temperature analysis
performed on spatial profiles  of the 24 and 70 $\mu$m data obtained near HD 34078 
show elevated dust temperatures ($\gtrsim$~90 K) in a bow shock ahead of the star, 
in the direction of stellar motion.  The bow shock scenario is
supported by the cometary morphological structure seen in the 8.0 and 
24 $\mu$m images of the stellar region.  We use this data to support the
two-component model for the HD 34078 sightline.  The excited H$_{2}$ 
is thought to be associated with the shock region, while the majority of the cold H$_{2}$ absorption 
and dust extinction  occurs in an intervening translucent cloud.

The measured AEF line strength were found to be anti-correlated with the intensity 
of the far-UV radiation field in the nebula, possibly due to increased
destruction of the aromatics.  The spectroscopic data also were 
used to constrain the relative ionization state of the AEF carrier
molecules. The ionization state was seen to increase with decreasing
density in one filament.
This ionization fraction was seen to be anticorrelated with the H$_{2}$ emission
line strength, consistent with enhanced AEF carrier ionization
with decreasing density at a given UV radiation field.
Evidence was presented for substructure in the 6.2 $\mu$m
AEF, suggestive of changing carrier size and/or composition as a function of
nebular environment.   Line ratios were used to 
derive the H$_{2}$ rotational temperature in two filaments, indicating an 
average rotational temperature of roughly 400 K, with evidence for additional 
non-uniform excitation in the filament nearer to the illuminating star.  These measurements 
were complemented by observations of the far-UV H$_{2}$
fluorescence spectrum, providing direct evidence for excitation by UV photons.
  
As new imaging and spectroscopic observations of PDRs become available in the
$Spitzer$ archive, opportunities for comparative studies with existing 
UV data are made possible.  Combining the UV and mid-IR observations
with near-IR observations of the rovibrational emission lines of H$_{2}$, 
an accounting of the energy redistribution 
following electronic excitation can be carried out in order to 
test models of H$_{2}$ fluorescence.  A panchromatic H$_{2}$ spectrum 
could be produced in a variety of PDRs, from pure reflection nebulae (NGC 2023) to star-forming
regions (Orion Bar) to \ion{H}{2} regions with harder radiation fields (planetary nebulae
such as M 27).  These data would allow models that span a wide range in 
excitation temperature, density, and radiation field to be compared with observation.
H$_{2}$ observations can be compared to the AEF and dust properties in these regions
in order to explore the relationship between molecular-phase PDR diagnostics and
local environmental characteristics.

\acknowledgments
It is a pleasure to acknowledge B-G Andersson for several
discussions about the infrared properties of dust grains.  
We thank an anonymous referee, whose comments greatly improved the
quality of this paper.
This work is based in part on observations made with the $Spitzer$ 
$Space$ $Telescope$, which is operated by the Jet Propulsion Laboratory, California Institute of 
Technology under a contract with NASA.  Support for this work was 
provided by JPL grant JPL1276492 to The Johns Hopkins University.  
The $FUSE$ data were obtained under the 
Guest Investigator Program of the NASA-CNES-CSA $FUSE$ mission, operated
by The Johns Hopkins University.



\begin{thebibliography}{77}
\expandafter\ifx\csname natexlab\endcsname\relax\def\natexlab#1{#1}\fi

\bibitem[{{Abel} {et~al.}(2005){Abel}, {Ferland}, {Shaw}, \& {van
  Hoof}}]{abel05}
{Abel}, N.~P., {Ferland}, G.~J., {Shaw}, G., \& {van Hoof}, P.~A.~M. 2005,
  \apjs, 161, 65

\bibitem[{{Allers} {et~al.}(2005){Allers}, {Jaffe}, {Lacy}, {Draine}, \&
  {Richter}}]{allers05}
{Allers}, K.~N., {Jaffe}, D.~T., {Lacy}, J.~H., {Draine}, B.~T., \& {Richter},
  M.~J. 2005, \apj, 630, 368

\bibitem[{{Armus} {et~al.}(2004){Armus}, {Charmandaris}, {Spoon}, {Houck},
  {Soifer}, {Brandl}, {Appleton}, {Teplitz}, {Higdon}, {Weedman}, {Devost},
  {Morris}, {Uchida}, {van Cleve}, {Barry}, {Sloan}, {Grillmair}, {Burgdorf},
  {Fajardo-Acosta}, {Ingalls}, {Higdon}, {Hao}, {Bernard-Salas}, {Herter},
  {Troeltzsch}, {Unruh}, \& {Winghart}}]{armus04}
{Armus}, L., {Charmandaris}, V., {Spoon}, H.~W.~W., {Houck}, J.~R., {Soifer},
  B.~T., {Brandl}, B.~R., {Appleton}, P.~N., {Teplitz}, H.~I., {Higdon},
  S.~J.~U., {Weedman}, D.~W., {Devost}, D., {Morris}, P.~W., {Uchida}, K.~I.,
  {van Cleve}, J., {Barry}, D.~J., {Sloan}, G.~C., {Grillmair}, C.~J.,
  {Burgdorf}, M.~J., {Fajardo-Acosta}, S.~B., {Ingalls}, J.~G., {Higdon}, J.,
  {Hao}, L., {Bernard-Salas}, J., {Herter}, T., {Troeltzsch}, J., {Unruh}, B.,
  \& {Winghart}, M. 2004, \apjs, 154, 178

\bibitem[{{Bagnuolo} {et~al.}(2001){Bagnuolo}, {Riddle}, {Gies}, \&
  {Barry}}]{bagnuolo01}
{Bagnuolo}, Jr., W.~G., {Riddle}, R.~L., {Gies}, D.~R., \& {Barry}, D.~J. 2001,
  \apj, 554, 362

\bibitem[{{Bakes} {et~al.}(2001){Bakes}, {Tielens}, \&
  {Bauschlicher}}]{bakes01}
{Bakes}, E.~L.~O., {Tielens}, A.~G.~G.~M., \& {Bauschlicher}, Jr., C.~W. 2001,
  \apj, 556, 501

\bibitem[{{Blaauw} \& {Morgan}(1954)}]{blaauw54}
{Blaauw}, A. \& {Morgan}, W.~W. 1954, \apj, 119, 625

\bibitem[{{Black} \& {van Dishoeck}(1987)}]{black87}
{Black}, J.~H. \& {van Dishoeck}, E.~F. 1987, \apj, 322, 412

\bibitem[{{Boiss{\'e}} {et~al.}(2005){Boiss{\'e}}, {Le Petit}, {Rollinde},
  {Roueff}, {Pineau des For{\^e}ts}, {Andersson}, {Gry}, \&
  {Felenbok}}]{boisse05}
{Boiss{\'e}}, P., {Le Petit}, F., {Rollinde}, E., {Roueff}, E., {Pineau des
  For{\^e}ts}, G., {Andersson}, B.-G., {Gry}, C., \& {Felenbok}, P. 2005, \aap,
  429, 509

\bibitem[{{Boulanger} {et~al.}(2000){Boulanger}, {Abergel}, {Cesarsky},
  {Bernard}, {Miville Desch{\^e}nes}, {Verstraete}, \& {Reach}}]{boulanger00}
{Boulanger}, F., {Abergel}, A., {Cesarsky}, D., {Bernard}, J.~P., {Miville
  Desch{\^e}nes}, M.~A., {Verstraete}, L., \& {Reach}, W.~T. 2000, in ESA
  SP-455: ISO Beyond Point Sources: Studies of Extended Infrared Emission, ed.
  R.~J. {Laureijs}, K.~{Leech}, \& M.~F. {Kessler}, 91--+

\bibitem[{{Bregman} \& {Temi}(2005)}]{bregman05}
{Bregman}, J. \& {Temi}, P. 2005, \apj, 621, 831

\bibitem[{{Bregman} {et~al.}(1989){Bregman}, {Allamandola}, {Witteborn},
  {Tielens}, \& {Geballe}}]{bregman89}
{Bregman}, J.~D., {Allamandola}, L.~J., {Witteborn}, F.~C., {Tielens},
  A.~G.~G.~M., \& {Geballe}, T.~R. 1989, \apj, 344, 791

\bibitem[{{Burgh} {et~al.}(2002){Burgh}, {McCandliss}, \& {Feldman}}]{burgh02}
{Burgh}, E.~B., {McCandliss}, S.~R., \& {Feldman}, P.~D. 2002, ApJ, 575, 240

\bibitem[{{Cardelli} {et~al.}(1989){Cardelli}, {Clayton}, \& {Mathis}}]{ccm}
{Cardelli}, J.~A., {Clayton}, G.~C., \& {Mathis}, J.~S. 1989, \apj, 345, 245

\bibitem[{{Cazaux} \& {Tielens}(2002)}]{cazaux02}
{Cazaux}, S. \& {Tielens}, A.~G.~G.~M. 2002, \apjl, 575, L29

\bibitem[{{Cazaux} \& {Tielens}(2004)}]{cazaux04}
---. 2004, \apj, 604, 222

\bibitem[{{Cesarsky} {et~al.}(2000){Cesarsky}, {Lequeux}, {Ryter}, \&
  {G{\'e}rin}}]{cesarsky00}
{Cesarsky}, D., {Lequeux}, J., {Ryter}, C., \& {G{\'e}rin}, M. 2000, \aap, 354,
  L87

\bibitem[{{Chan} {et~al.}(2001){Chan}, {Roellig}, {Onaka}, {Mizutani},
  {Okumura}, {Yamamura}, {Tanab{\'e}}, {Shibai}, {Nakagawa}, \&
  {Okuda}}]{chan01}
{Chan}, K.-W., {Roellig}, T.~L., {Onaka}, T., {Mizutani}, M., {Okumura}, K.,
  {Yamamura}, I., {Tanab{\'e}}, T., {Shibai}, H., {Nakagawa}, T., \& {Okuda},
  H. 2001, \apj, 546, 273

\bibitem[{{Cohen} {et~al.}(1985){Cohen}, {Tielens}, \& {Allamandola}}]{cohen85}
{Cohen}, M., {Tielens}, A.~G.~G.~M., \& {Allamandola}, L.~J. 1985, \apjl, 299,
  L93

\bibitem[{{Draine}(2003)}]{draine03}
{Draine}, B.~T. 2003, \apj, 598, 1017

\bibitem[{{Draine} \& {Li}(2001)}]{draine01}
{Draine}, B.~T. \& {Li}, A. 2001, \apj, 551, 807

\bibitem[{{Draine} {et~al.}(1983){Draine}, {Roberge}, \& {Dalgarno}}]{draine83}
{Draine}, B.~T., {Roberge}, W.~G., \& {Dalgarno}, A. 1983, \apj, 264, 485

\bibitem[{{Fazio} {et~al.}(2004){Fazio}, {Hora}, {Allen}, {Ashby}, {Barmby},
  {Deutsch}, {Huang}, {Kleiner}, {Marengo}, {Megeath}, {Melnick}, {Pahre},
  {Patten}, {Polizotti}, {Smith}, {Taylor}, {Wang}, {Willner}, {Hoffmann},
  {Pipher}, {Forrest}, {McMurty}, {McCreight}, {McKelvey}, {McMurray}, {Koch},
  {Moseley}, {Arendt}, {Mentzell}, {Marx}, {Losch}, {Mayman}, {Eichhorn},
  {Krebs}, {Jhabvala}, {Gezari}, {Fixsen}, {Flores}, {Shakoorzadeh}, {Jungo},
  {Hakun}, {Workman}, {Karpati}, {Kichak}, {Whitley}, {Mann}, {Tollestrup},
  {Eisenhardt}, {Stern}, {Gorjian}, {Bhattacharya}, {Carey}, {Nelson},
  {Glaccum}, {Lacy}, {Lowrance}, {Laine}, {Reach}, {Stauffer}, {Surace},
  {Wilson}, {Wright}, {Hoffman}, {Domingo}, \& {Cohen}}]{irac04}
{Fazio}, G.~G., {Hora}, J.~L., {Allen}, L.~E., {Ashby}, M.~L.~N., {Barmby}, P.,
  {Deutsch}, L.~K., {Huang}, J.-S., {Kleiner}, S., {Marengo}, M., {Megeath},
  S.~T., {Melnick}, G.~J., {Pahre}, M.~A., {Patten}, B.~M., {Polizotti}, J.,
  {Smith}, H.~A., {Taylor}, R.~S., {Wang}, Z., {Willner}, S.~P., {Hoffmann},
  W.~F., {Pipher}, J.~L., {Forrest}, W.~J., {McMurty}, C.~W., {McCreight},
  C.~R., {McKelvey}, M.~E., {McMurray}, R.~E., {Koch}, D.~G., {Moseley}, S.~H.,
  {Arendt}, R.~G., {Mentzell}, J.~E., {Marx}, C.~T., {Losch}, P., {Mayman}, P.,
  {Eichhorn}, W., {Krebs}, D., {Jhabvala}, M., {Gezari}, D.~Y., {Fixsen},
  D.~J., {Flores}, J., {Shakoorzadeh}, K., {Jungo}, R., {Hakun}, C., {Workman},
  L., {Karpati}, G., {Kichak}, R., {Whitley}, R., {Mann}, S., {Tollestrup},
  E.~V., {Eisenhardt}, P., {Stern}, D., {Gorjian}, V., {Bhattacharya}, B.,
  {Carey}, S., {Nelson}, B.~O., {Glaccum}, W.~J., {Lacy}, M., {Lowrance},
  P.~J., {Laine}, S., {Reach}, W.~T., {Stauffer}, J.~A., {Surace}, J.~A.,
  {Wilson}, G., {Wright}, E.~L., {Hoffman}, A., {Domingo}, G., \& {Cohen}, M.
  2004, \apjs, 154, 10

\bibitem[{{France} {et~al.}(2005){France}, {Andersson}, {McCandliss}, \&
  {Feldman}}]{france05a}
{France}, K., {Andersson}, B.-G., {McCandliss}, S.~R., \& {Feldman}, P.~D.
  2005, \apj, 628, 750

\bibitem[{{France} \& {McCandliss}(2005)}]{france05b}
{France}, K. \& {McCandliss}, S.~R. 2005, \apjl, 629, L97

\bibitem[{{France} {et~al.}(2004){France}, {McCandliss}, {Burgh}, \&
  {Feldman}}]{france04}
{France}, K., {McCandliss}, S.~R., {Burgh}, E.~B., \& {Feldman}, P.~D. 2004,
  \apj, 616, 257

\bibitem[{{Gordon} {et~al.}(2006){Gordon}, {Englebracht}, {Smith}, {Rieke}, \&
  {Misselt}}]{gordon06}
{Gordon}, K., {Englebracht}, C., {Smith}, J., {Rieke}, G., \& {Misselt}, K.
  2006, astro-ph, 0605544

\bibitem[{{Habart} {et~al.}(2004){Habart}, {Boulanger}, {Verstraete},
  {Walmsley}, \& {Pineau des For{\^e}ts}}]{habart04}
{Habart}, E., {Boulanger}, F., {Verstraete}, L., {Walmsley}, C.~M., \& {Pineau
  des For{\^e}ts}, G. 2004, \aap, 414, 531

\bibitem[{{Heitsch} {et~al.}(2006){Heitsch}, {Whittney}, {Indebtown}, {Meade},
  {Babler}, \& {Churchwell}}]{heitsch06}
{Heitsch}, R., {Whittney}, B.~A., {Indebtown}, R., {Meade}, M.~R., {Babler},
  B.~L., \& {Churchwell}, E. 2006, \apj, astroph0607318

\bibitem[{{Herbig}(1958)}]{herbig58}
{Herbig}, G.~H. 1958, \pasp, 70, 468

\bibitem[{{Herbig}(1999)}]{herbig99}
---. 1999, \pasp, 111, 809

\bibitem[{{Hollenbach} \& {Tielens}(1997)}]{pdrs}
{Hollenbach}, D. \& {Tielens}, A. 1997, Annual Reviews of Astronomy and
  Astrophysics, 35, 179

\bibitem[{{Hony} {et~al.}(2001){Hony}, {Van Kerckhoven}, {Peeters}, {Tielens},
  {Hudgins}, \& {Allamandola}}]{hony01}
{Hony}, S., {Van Kerckhoven}, C., {Peeters}, E., {Tielens}, A.~G.~G.~M.,
  {Hudgins}, D.~M., \& {Allamandola}, L.~J. 2001, \aap, 370, 1030

\bibitem[{{Hora} {et~al.}(2006){Hora}, {Latter}, {Smith}, \&
  {Marengo}}]{hora06}
{Hora}, J., {Latter}, W.~B., {Smith}, H., \& {Marengo}, M. 2006, \apj,
  astroph0607541

\bibitem[{{Houck} {et~al.}(2004){Houck}, {Roellig}, {van Cleve}, {Forrest},
  {Herter}, {Lawrence}, {Matthews}, {Reitsema}, {Soifer}, {Watson}, {Weedman},
  {Huisjen}, {Troeltzsch}, {Barry}, {Bernard-Salas}, {Blacken}, {Brandl},
  {Charmandaris}, {Devost}, {Gull}, {Hall}, {Henderson}, {Higdon}, {Pirger},
  {Schoenwald}, {Sloan}, {Uchida}, {Appleton}, {Armus}, {Burgdorf},
  {Fajardo-Acosta}, {Grillmair}, {Ingalls}, {Morris}, \& {Teplitz}}]{irs04}
{Houck}, J.~R., {Roellig}, T.~L., {van Cleve}, J., {Forrest}, W.~J., {Herter},
  T., {Lawrence}, C.~R., {Matthews}, K., {Reitsema}, H.~J., {Soifer}, B.~T.,
  {Watson}, D.~M., {Weedman}, D., {Huisjen}, M., {Troeltzsch}, J., {Barry},
  D.~J., {Bernard-Salas}, J., {Blacken}, C.~E., {Brandl}, B.~R.,
  {Charmandaris}, V., {Devost}, D., {Gull}, G.~E., {Hall}, P., {Henderson},
  C.~P., {Higdon}, S.~J.~U., {Pirger}, B.~E., {Schoenwald}, J., {Sloan}, G.~C.,
  {Uchida}, K.~I., {Appleton}, P.~N., {Armus}, L., {Burgdorf}, M.~J.,
  {Fajardo-Acosta}, S.~B., {Grillmair}, C.~J., {Ingalls}, J.~G., {Morris},
  P.~W., \& {Teplitz}, H.~I. 2004, \apjs, 154, 18

\bibitem[{{Hudgins} {et~al.}(2005){Hudgins}, {Bauschlicher}, \&
  {Allamandola}}]{hudgins05}
{Hudgins}, D.~M., {Bauschlicher}, Jr., C.~W., \& {Allamandola}, L.~J. 2005,
  \apj, 632, 316

\bibitem[{{Huggins} {et~al.}(2002){Huggins}, {Forveille}, {Bachiller}, {Cox},
  {Ageorges}, \& {Walsh}}]{huggins02}
{Huggins}, P.~J., {Forveille}, T., {Bachiller}, R., {Cox}, P., {Ageorges}, N.,
  \& {Walsh}, J.~R. 2002, \apjl, 573, L55

\bibitem[{{Hurwitz}(1998)}]{hurwitz98}
{Hurwitz}, M. 1998, \apjl, 500, L67+

\bibitem[{{Huthoff} \& {Kaper}(2002)}]{huthoff02}
{Huthoff}, F. \& {Kaper}, L. 2002, \aap, 383, 999

\bibitem[{{Jansen} {et~al.}(1994){Jansen}, {van Dishoeck}, \&
  {Black}}]{jansen94}
{Jansen}, D.~J., {van Dishoeck}, E.~F., \& {Black}, J.~H. 1994, \aap, 282, 605

\bibitem[{{Jansen} {et~al.}(1995){Jansen}, {van Dishoeck}, {Black}, {Spaans},
  \& {Sosin}}]{jansen95}
{Jansen}, D.~J., {van Dishoeck}, E.~F., {Black}, J.~H., {Spaans}, M., \&
  {Sosin}, C. 1995, \aap, 302, 223

\bibitem[{{Kastner} {et~al.}(1996){Kastner}, {Weintraub}, {Gatley}, {Merrill},
  \& {Probst}}]{kastner96}
{Kastner}, J.~H., {Weintraub}, D.~A., {Gatley}, I., {Merrill}, K.~M., \&
  {Probst}, R.~G. 1996, \apj, 462, 777

\bibitem[{{Kristensen} {et~al.}(2003){Kristensen}, {Gustafsson}, {Field},
  {Callejo}, {Lemaire}, {Vannier}, \& {Pineau des For{\^e}ts}}]{kristensen03}
{Kristensen}, L.~E., {Gustafsson}, M., {Field}, D., {Callejo}, G., {Lemaire},
  J.~L., {Vannier}, L., \& {Pineau des For{\^e}ts}, G. 2003, \aap, 412, 727

\bibitem[{{Li} \& {Draine}(2001)}]{li01}
{Li}, A. \& {Draine}, B.~T. 2001, \apj, 554, 778

\bibitem[{{Lis} {et~al.}(1998){Lis}, {Serabyn}, {Keene}, {Dowell}, {Benford},
  {Phillips}, {Hunter}, \& {Wang}}]{lis98}
{Lis}, D.~C., {Serabyn}, E., {Keene}, J., {Dowell}, C.~D., {Benford}, D.~J.,
  {Phillips}, T.~G., {Hunter}, T.~R., \& {Wang}, N. 1998, \apj, 509, 299

\bibitem[{{Luhman} \& {Jaffe}(1996)}]{luhman96}
{Luhman}, M.~L. \& {Jaffe}, D.~T. 1996, \apj, 463, 191

\bibitem[{{Luhman} {et~al.}(1997){Luhman}, {Luhman}, {Benedict}, {Jaffe}, \&
  {Fischer}}]{luhman97}
{Luhman}, M.~L., {Luhman}, K.~L., {Benedict}, T., {Jaffe}, D.~T., \& {Fischer},
  J. 1997, \apjl, 480, L133+

\bibitem[{{Lupu} {et~al.}(2006){Lupu}, {France}, \& {McCandliss}}]{lupu06}
{Lupu}, R.~E., {France}, K., \& {McCandliss}, S.~R. 2006, \apj, 644, 981

\bibitem[{{Mac Low} {et~al.}(1991){Mac Low}, {van Buren}, {Wood}, \&
  {Churchwell}}]{maclow91}
{Mac Low}, M.-M., {van Buren}, D., {Wood}, D.~O.~S., \& {Churchwell}, E. 1991,
  \apj, 369, 395

\bibitem[{{Mandy} \& {Martin}(1993)}]{mandy93}
{Mandy}, M.~E. \& {Martin}, P.~G. 1993, \apjs, 86, 199

\bibitem[{{Martini} {et~al.}(1999){Martini}, {Sellgren}, \&
  {DePoy}}]{martini99}
{Martini}, P., {Sellgren}, K., \& {DePoy}, D.~L. 1999, \apj, 526, 772

\bibitem[{{Martins} {et~al.}(2005){Martins}, {Schaerer}, {Hillier},
  {Meynadier}, {Heydari-Malayeri}, \& {Walborn}}]{martins05}
{Martins}, F., {Schaerer}, D., {Hillier}, D.~J., {Meynadier}, F.,
  {Heydari-Malayeri}, M., \& {Walborn}, N.~R. 2005, \aap, 441, 735

\bibitem[{{Moos}(2000)}]{moos00}
{Moos}, H.~W. e.~a. 2000, \apjl, 538, L1

\bibitem[{{Noriega-Crespo} {et~al.}(1997){Noriega-Crespo}, {van Buren}, \&
  {Dgani}}]{noriega97}
{Noriega-Crespo}, A., {van Buren}, D., \& {Dgani}, R. 1997, \aj, 113, 780

\bibitem[{{Peeters} {et~al.}(2002){Peeters}, {Hony}, {Van Kerckhoven},
  {Tielens}, {Allamandola}, {Hudgins}, \& {Bauschlicher}}]{peeters02}
{Peeters}, E., {Hony}, S., {Van Kerckhoven}, C., {Tielens}, A.~G.~G.~M.,
  {Allamandola}, L.~J., {Hudgins}, D.~M., \& {Bauschlicher}, C.~W. 2002, \aap,
  390, 1089

\bibitem[{{Peeters} {et~al.}(2004){Peeters}, {Mattioda}, {Hudgins}, \&
  {Allamandola}}]{peeters04}
{Peeters}, E., {Mattioda}, A.~L., {Hudgins}, D.~M., \& {Allamandola}, L.~J.
  2004, \apjl, 617, L65

\bibitem[{{Pellerin} {et~al.}(2002){Pellerin}, {Fullerton}, {Robert}, {Howk},
  {Hutchings}, {Walborn}, {Bianchi}, {Crowther}, \& {Sonneborn}}]{pellerin02}
{Pellerin}, A., {Fullerton}, A.~W., {Robert}, C., {Howk}, J.~C., {Hutchings},
  J.~B., {Walborn}, N.~R., {Bianchi}, L., {Crowther}, P.~A., \& {Sonneborn}, G.
  2002, \apjs, 143, 159

\bibitem[{{Poglitsch} {et~al.}(2004){Poglitsch}, {Waelkens}, {Bauer}, {Cepa},
  {Henning}, {van Hoof}, {Katterloher}, {Kerschbaum}, {Lemke}, {Renotte},
  {Rodriguez}, {Royer}, \& {Saraceno}}]{pacs04}
{Poglitsch}, A., {Waelkens}, C., {Bauer}, O.~H., {Cepa}, J., {Henning}, T.~F.,
  {van Hoof}, C., {Katterloher}, R., {Kerschbaum}, F., {Lemke}, D., {Renotte},
  E., {Rodriguez}, L., {Royer}, P., \& {Saraceno}, P. 2004, in Microwave and
  Terahertz Photonics. Edited by Stohr, Andreas; Jager, Dieter; Iezekiel,
  Stavros. Proceedings of the SPIE, Volume 5487, pp. 425-436 (2004)., ed. J.~C.
  {Mather}, 425--436

\bibitem[{{Rieke} {et~al.}(2004){Rieke}, {Young}, {Engelbracht}, {Kelly},
  {Low}, {Haller}, {Beeman}, {Gordon}, {Stansberry}, {Misselt}, {Cadien},
  {Morrison}, {Rivlis}, {Latter}, {Noriega-Crespo}, {Padgett}, {Stapelfeldt},
  {Hines}, {Egami}, {Muzerolle}, {Alonso-Herrero}, {Blaylock}, {Dole}, {Hinz},
  {Le Floc'h}, {Papovich}, {P{\'e}rez-Gonz{\'a}lez}, {Smith}, {Su}, {Bennett},
  {Frayer}, {Henderson}, {Lu}, {Masci}, {Pesenson}, {Rebull}, {Rho}, {Keene},
  {Stolovy}, {Wachter}, {Wheaton}, {Werner}, \& {Richards}}]{mips04}
{Rieke}, G.~H., {Young}, E.~T., {Engelbracht}, C.~W., {Kelly}, D.~M., {Low},
  F.~J., {Haller}, E.~E., {Beeman}, J.~W., {Gordon}, K.~D., {Stansberry},
  J.~A., {Misselt}, K.~A., {Cadien}, J., {Morrison}, J.~E., {Rivlis}, G.,
  {Latter}, W.~B., {Noriega-Crespo}, A., {Padgett}, D.~L., {Stapelfeldt},
  K.~R., {Hines}, D.~C., {Egami}, E., {Muzerolle}, J., {Alonso-Herrero}, A.,
  {Blaylock}, M., {Dole}, H., {Hinz}, J.~L., {Le Floc'h}, E., {Papovich}, C.,
  {P{\'e}rez-Gonz{\'a}lez}, P.~G., {Smith}, P.~S., {Su}, K.~Y.~L., {Bennett},
  L., {Frayer}, D.~T., {Henderson}, D., {Lu}, N., {Masci}, F., {Pesenson}, M.,
  {Rebull}, L., {Rho}, J., {Keene}, J., {Stolovy}, S., {Wachter}, S.,
  {Wheaton}, W., {Werner}, M.~W., \& {Richards}, P.~L. 2004, \apjs, 154, 25

\bibitem[{{Rollinde} {et~al.}(2003){Rollinde}, {Boiss{\' e}}, {Federman}, \&
  {Pan}}]{rollinde03}
{Rollinde}, E., {Boiss{\' e}}, P., {Federman}, S.~R., \& {Pan}, K. 2003, \aap,
  401, 215

\bibitem[{{Rosenthal} {et~al.}(2000){Rosenthal}, {Bertoldi}, \&
  {Drapatz}}]{rosenthal00}
{Rosenthal}, D., {Bertoldi}, F., \& {Drapatz}, S. 2000, \aap, 356, 705

\bibitem[{{Schnee} {et~al.}(2005){Schnee}, {Ridge}, {Goodman}, \&
  {Li}}]{schnee05}
{Schnee}, S.~L., {Ridge}, N.~A., {Goodman}, A.~A., \& {Li}, J.~G. 2005, \apj,
  634, 442

\bibitem[{{Schutte} {et~al.}(1993){Schutte}, {Tielens}, \&
  {Allamandola}}]{schutte93}
{Schutte}, W.~A., {Tielens}, A.~G.~G.~M., \& {Allamandola}, L.~J. 1993, \apj,
  415, 397

\bibitem[{{Shaw} {et~al.}(2005){Shaw}, {Ferland}, {Abel}, {Stancil}, \& {van
  Hoof}}]{shaw05}
{Shaw}, G., {Ferland}, G.~J., {Abel}, N.~P., {Stancil}, P.~C., \& {van Hoof},
  P.~A.~M. 2005, \apj, 624, 794

\bibitem[{{Shull} \& {Beckwith}(1982)}]{shull82}
{Shull}, M. \& {Beckwith}, S. 1982, Annual Reviews of Astronomy and
  Astrophysics, 30, 163

\bibitem[{{Sodroski} {et~al.}(1987){Sodroski}, {Dwek}, {Hauser}, \&
  {Kerr}}]{sodroski87}
{Sodroski}, T.~J., {Dwek}, E., {Hauser}, M.~G., \& {Kerr}, F.~J. 1987, \apj,
  322, 101

\bibitem[{{Speck} {et~al.}(2003){Speck}, {Meixner}, {Jacoby}, \&
  {Knezek}}]{speck03}
{Speck}, A.~K., {Meixner}, M., {Jacoby}, G.~H., \& {Knezek}, P.~M. 2003, \pasp,
  115, 170

\bibitem[{{Stecher} \& {Williams}(1967)}]{stecher67}
{Stecher}, T.~P. \& {Williams}, D.~A. 1967, \apjl, 149, L29+

\bibitem[{{Takami} {et~al.}(2000){Takami}, {Usuda}, {Sugai}, {Kawabata},
  {Suto}, \& {Tanaka}}]{takami00}
{Takami}, M., {Usuda}, T., {Sugai}, H., {Kawabata}, H., {Suto}, H., \&
  {Tanaka}, M. 2000, \apj, 529, 268

\bibitem[{{Thi} {et~al.}(1999){Thi}, {van Dishoeck}, {Black}, {Jansen},
  {Evans}, \& {Jaffe}}]{thi99}
{Thi}, W.~F., {van Dishoeck}, E.~F., {Black}, J.~H., {Jansen}, D.~J., {Evans},
  N.~J., \& {Jaffe}, D.~T. 1999, in ESA SP-427: The Universe as Seen by ISO,
  ed. P.~{Cox} \& M.~{Kessler}, 529--+

\bibitem[{{Uchida} {et~al.}(2000){Uchida}, {Sellgren}, {Werner}, \&
  {Houdashelt}}]{uchida00}
{Uchida}, K.~I., {Sellgren}, K., {Werner}, M.~W., \& {Houdashelt}, M.~L. 2000,
  \apj, 530, 817

\bibitem[{{Valentijn} \& {van der Werf}(1999)}]{valentijn99}
{Valentijn}, E.~A. \& {van der Werf}, P.~P. 1999, \apjl, 522, L29

\bibitem[{{van Buren} {et~al.}(1990){van Buren}, {Mac Low}, {Wood}, \&
  {Churchwell}}]{vanburen90}
{van Buren}, D., {Mac Low}, M.-M., {Wood}, D.~O.~S., \& {Churchwell}, E. 1990,
  \apj, 353, 570

\bibitem[{{van Buren} \& {McCray}(1988)}]{vanburen88}
{van Buren}, D. \& {McCray}, R. 1988, \apjl, 329, L93

\bibitem[{{van Buren} {et~al.}(1995){van Buren}, {Noriega-Crespo}, \&
  {Dgani}}]{vanburen95}
{van Buren}, D., {Noriega-Crespo}, A., \& {Dgani}, R. 1995, \aj, 110, 2914

\bibitem[{{van Dishoeck}(2004)}]{vand04}
{van Dishoeck}, E.~F. 2004, \araa, 42, 119

\bibitem[{{Verstraete} {et~al.}(2001){Verstraete}, {Pech}, {Moutou},
  {Sellgren}, {Wright}, {Giard}, {L{\'e}ger}, {Timmermann}, \&
  {Drapatz}}]{verstraete01}
{Verstraete}, L., {Pech}, C., {Moutou}, C., {Sellgren}, K., {Wright}, C.~M.,
  {Giard}, M., {L{\'e}ger}, A., {Timmermann}, R., \& {Drapatz}, S. 2001, \aap,
  372, 981

\bibitem[{{Verstraete} {et~al.}(1996){Verstraete}, {Puget}, {Falgarone},
  {Drapatz}, {Wright}, \& {Timmermann}}]{verstraete96}
{Verstraete}, L., {Puget}, J.~L., {Falgarone}, E., {Drapatz}, S., {Wright},
  C.~M., \& {Timmermann}, R. 1996, \aap, 315, L337

\bibitem[{{Werner} {et~al.}(2004{\natexlab{a}}){Werner}, {Roellig}, {Low},
  {Rieke}, {Rieke}, {Hoffmann}, {Young}, {Houck}, {Brandl}, {Fazio}, {Hora},
  {Gehrz}, {Helou}, {Soifer}, {Stauffer}, {Keene}, {Eisenhardt}, {Gallagher},
  {Gautier}, {Irace}, {Lawrence}, {Simmons}, {Van Cleve}, {Jura}, {Wright}, \&
  {Cruikshank}}]{werner04}
{Werner}, M.~W., {Roellig}, T.~L., {Low}, F.~J., {Rieke}, G.~H., {Rieke}, M.,
  {Hoffmann}, W.~F., {Young}, E., {Houck}, J.~R., {Brandl}, B., {Fazio}, G.~G.,
  {Hora}, J.~L., {Gehrz}, R.~D., {Helou}, G., {Soifer}, B.~T., {Stauffer}, J.,
  {Keene}, J., {Eisenhardt}, P., {Gallagher}, D., {Gautier}, T.~N., {Irace},
  W., {Lawrence}, C.~R., {Simmons}, L., {Van Cleve}, J.~E., {Jura}, M.,
  {Wright}, E.~L., \& {Cruikshank}, D.~P. 2004{\natexlab{a}}, \apjs, 154, 1

\bibitem[{{Werner} {et~al.}(2004{\natexlab{b}}){Werner}, {Uchida}, {Sellgren},
  {Marengo}, {Gordon}, {Morris}, {Houck}, \& {Stansberry}}]{werner04b}
{Werner}, M.~W., {Uchida}, K.~I., {Sellgren}, K., {Marengo}, M., {Gordon},
  K.~D., {Morris}, P.~W., {Houck}, J.~R., \& {Stansberry}, J.~A.
  2004{\natexlab{b}}, \apjs, 154, 309

\bibitem[{{Witt} {et~al.}(1993){Witt}, {Petersohn}, {Holberg}, {Murthy},
  {Dring}, \& {Henry}}]{witt93}
{Witt}, A.~N., {Petersohn}, J.~K., {Holberg}, J.~B., {Murthy}, J., {Dring}, A.,
  \& {Henry}, R.~C. 1993, ApJ, 410, 714

\bibitem[{{Witt} {et~al.}(1989){Witt}, {Stecher}, {Boroson}, \&
  {Bohlin}}]{witt89}
{Witt}, A.~N., {Stecher}, T.~P., {Boroson}, T.~A., \& {Bohlin}, R.~C. 1989,
  \apjl, 336, L21

\bibitem[{{Wolniewicz} {et~al.}(1998){Wolniewicz}, {Simbotin}, \&
  {Dalgarno}}]{avals98}
{Wolniewicz}, L., {Simbotin}, I., \& {Dalgarno}, A. 1998, \apjs, 115, 293

\end{thebibliography}
\end{document}